# Pronounced scale-dependent charge carrier density in graphene quantum Hall devices

Ziqiang Kong[1,2]#, Yu Feng[1,2]#, Han Gao[3]#, Ru Sun[1,2]#, Jian Feng[5], Chengxin Jiang[1,2], Chenxi Liu[1,2], Huishan Wang[6], Yu Zhang[1,2], Junchi Song[1,2], Xuanzheng Hao[1,2], Ziceng Zhang[1,2], Yuteng Ma[7,8], Shengda Gao[9], Ren Zhu[10], Qandeel Noor[11], Ghulam Ali[11], Yumeng Yang[4], Guanghui Yu[1,2], Shujie Tang[1,2], Zhongkai Liu[3,*], Haomin Wang[1,2,*]

[1] State Key Laboratory of Materials for Integrated Circuits, Shanghai Institute of Microsystem and Information Technology, Chinese Academy of Sciences, Shanghai 200050, China

[2] Center of Materials Science and Optoelectronics Engineering, University of Chinese Academy of Sciences, Beijing 100049, China

[3] School of Physical Science and Technology, ShanghaiTech Laboratory for Topological Physics, ShanghaiTech University, Shanghai 201210, China

[4] School of Information Science and Technology, ShanghaiTech University, Shanghai 201210, China

[5] Shanghai Institute of Measurement and Testing Technology, Shanghai Key Laboratory of On-line Testing and Control Technology, Shanghai 201203, China

[6] Key Laboratory of Optoelectronic Material and Device, Department of Physics, Shanghai Normal University, Shanghai 200233, China

[7] Department of Electrical and Computer Engineering, National University of Singapore, Singapore 117583, Singapore

[8] National University of Singapore (Chong Qing) Research Institute, Chongqing Liang Jiang New Area, Chongqing 401123, China

[9] School of Advanced Technology, Xi'an Jiaotong-Liverpool University, Suzhou 215123, China

[10] Oxford Instruments, Shanghai 200030, China

[11] U.S.-Pakistan Center for Advanced Studies in Energy, National University of Sciences and Technology (NUST), H-12, Islamabad, Pakistan

*Corresponding author. E-mail: hmwang@mail.sim.ac.cn and liuzhk@shanghaitech.edu.cn

# These authors contributed equally to this work.

## Highlights

- Revealed for the first time the dependence of carrier density and Fermi level on channel width in SiC-grown graphene Hall devices.
- Combined Fermi velocity measurements and ARPES to demonstrate that scale-dependent carrier density originates from band structure modifications and electron-electron interactions in graphene.
- Optimized channel width (~360 μm) of graphene quantum resistance array via machine learning with limited data, achieving optimal balance between value uncertainty and on-chip integration density.

## Graphical Abstract

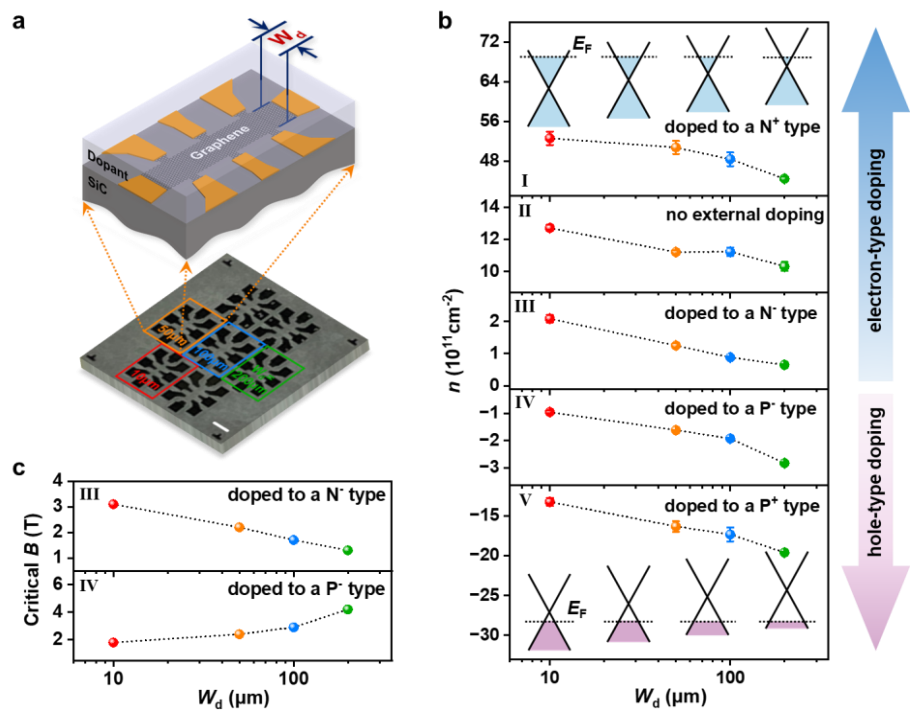

**Abstract**

The miniaturization of quantum Hall resistance standards (QHRS) using epitaxial graphene on silicon carbide (SiC) necessitates understanding how device dimensions impact performance. This study reveals a pronounced scale-dependent carrier density in graphene Hall devices: under electron doping, carrier density decreases with increasing channel width ($W_d$), while the opposite occurs under hole doping. This phenomenon, most significant for $W_d \leq 400$ μm, directly influences the onset of magnetic field required for quantization. Fermi velocity measurements and angle-resolved photoemission spectroscopy (ARPES) analysis indicate that band structure modifications and electron-electron interactions underlie this size dependence. Utilizing machine learning with limited data, we optimized the device geometry, identifying a channel width of ~360 μm as the optimal balance between resistance uncertainty and on-chip integration density. This work provides key insights for designing high-performance, miniaturized graphene-based QHRS arrays.

# 1. Introduction

Quantum Hall resistance standards (QHRS) based on epitaxial graphene on SiC have gained attention since the von Klitzing constant ($R_K = h/e^2$) was established as a universal reference for electrical resistance in the 2019 SI redefinition [1-3]. QHRS are primarily fabricated using graphene grown on SiC via chemical vapor deposition (CVD) [4-6] or silicon sublimation growth [7,8]. Graphene Hall devices enable QHRS to reach the $R_K/2$ plateau at lower magnetic fields with a reduced carrier concentration compared to other two-dimensional electron gas (2DEG) materials. To achieve this, doping methods such as adsorption (*e.g.*, oxygen [9], fluorinated fullerene [10], acid doping [5,11]), corona discharge [12,13], photochemical coatings [14,15] and stable dopants such as $Cr(CO)_3$ [16,17] and $F_4TCNQ$ [18-23] are used. QHRS with resistance values different from $R_K/2$ facilitate metrological applications, with widely achieved values of 109 Ω [24], 129 Ω [25], 1 kΩ [26], 8.604 kΩ [27,28] and 129 kΩ [29] using superconducting Nb-based interconnections. Expanding QHRS in arbitrary resistance values requires integrating more Hall devices on a single chip, prompting key questions on the minimum feasible device size and whether differently sized devices can coexist in one array without affecting performance. Despite its significance, the impact of device size on QHRS remains underexplored.

Literature suggests that wider graphene quantum Hall devices seem to more easily achieve the quantized resistance plateau at lower magnetic fields [1,2,4,11,13-28]. In this study, we explore this further using SiC-grown graphene Hall devices of varying widths. Doping modulation reveals that carrier density decreases with channel size under electron doping, but increases under hole doping. This trend, reflected in a downward Fermi level shift for narrower channels (especially below 400 μm), is attributed to subtle band structure modifications detected via Shubnikov–de Haas (SdH) oscillations and ARPES. To minimize operational magnetic fields and enhance integration—thus reducing costs—we employed machine learning to optimize the channel width. A width of ~360 μm was found to optimally balance uncertainty and on-chip integration density. Using this, we fabricated an $R_K/3$ array with $10^{-8}$-level uncertainty, advancing miniaturized, low-cost quantum resistance standards.

## 2. Results and Discussion

Fig. 1a presents a photograph and schematic diagram of graphene Hall devices with varying channel widths ($W_d$) on the same SiC substrate, ranging from 10 to 200 μm (fabrication processes are detailed in the Experimental Section). We spin-coated a dopant layer onto the graphene Hall devices and adjusted the doping condition by varying the dopant type and dopant concentration. Then, low-temperature magneto-transport measurements were conducted on the graphene Hall devices under different doping conditions, including $N^+$ type (I), no external doping (II), $N^-$ type (III), $P^-$ type (IV) and $P^+$ type (V) corresponding to Fig. 1b. The detailed low-temperature magneto-transport measurement results under these doping conditions are shown in Fig. S3–S7. Based on the measured curves of longitudinal resistance ($R_{xx}$) and Hall resistance ($R_{xy}$) as a function of magnetic field for devices with different channel widths, and using the formula $n = \frac{\sigma_{xx}}{\mu \cdot e} = -\frac{B}{R_{xy} \cdot e}$, a scatter plot of the relationship between carrier density ($n$) and $W_d$ can be obtained, as shown in Fig. 1b. Here, $\sigma_{xx}$ represents longitudinal conductivity, $\mu$ is carrier mobility, $e$ is electron charge, and $B$ is magnetic field strength [30]. It is worth noting that, in order to ensure consistent experimental conditions, devices with different channel widths on the same SiC substrate were measured simultaneously under each doping condition. Furthermore, as shown in Fig. S2, the carrier density of the doped graphene Hall devices demonstrates good temporal stability.

As shown in Fig. 1b, carrier density of graphene decreases with increasing $W_d$ in electron-type cases I, II, and III, whereas in hole-type cases IV and V, carrier density increases. Both trends indicate that the Fermi level ($E_F$) shifts downward as $W_d$ increases. Additionally, the insets in Fig. 1b ( I and V) provide schematic diagrams illustrating how $E_F$ of graphene decreases with increasing $W_d$ for both electron-type and hole-type doping. Fig. 1c illustrates the dependence of the onset of magnetic field ($B$) for entering quantized Hall plateaus at $h/2e^2$ on $W_d$ for both $N^-$ type (III) and $P^-$ type (IV) doping conditions. The detailed measurements of the onset of magnetic field are provided in Fig. S8. Under the $N^-$ type doping condition, the onset of $B$ decreases as $W_d$ increases. In contrast, under the $P^-$ type doping condition, the onset of $B$ increases with $W_d$. These trends result from the variation in carrier density under the corresponding doping conditions.

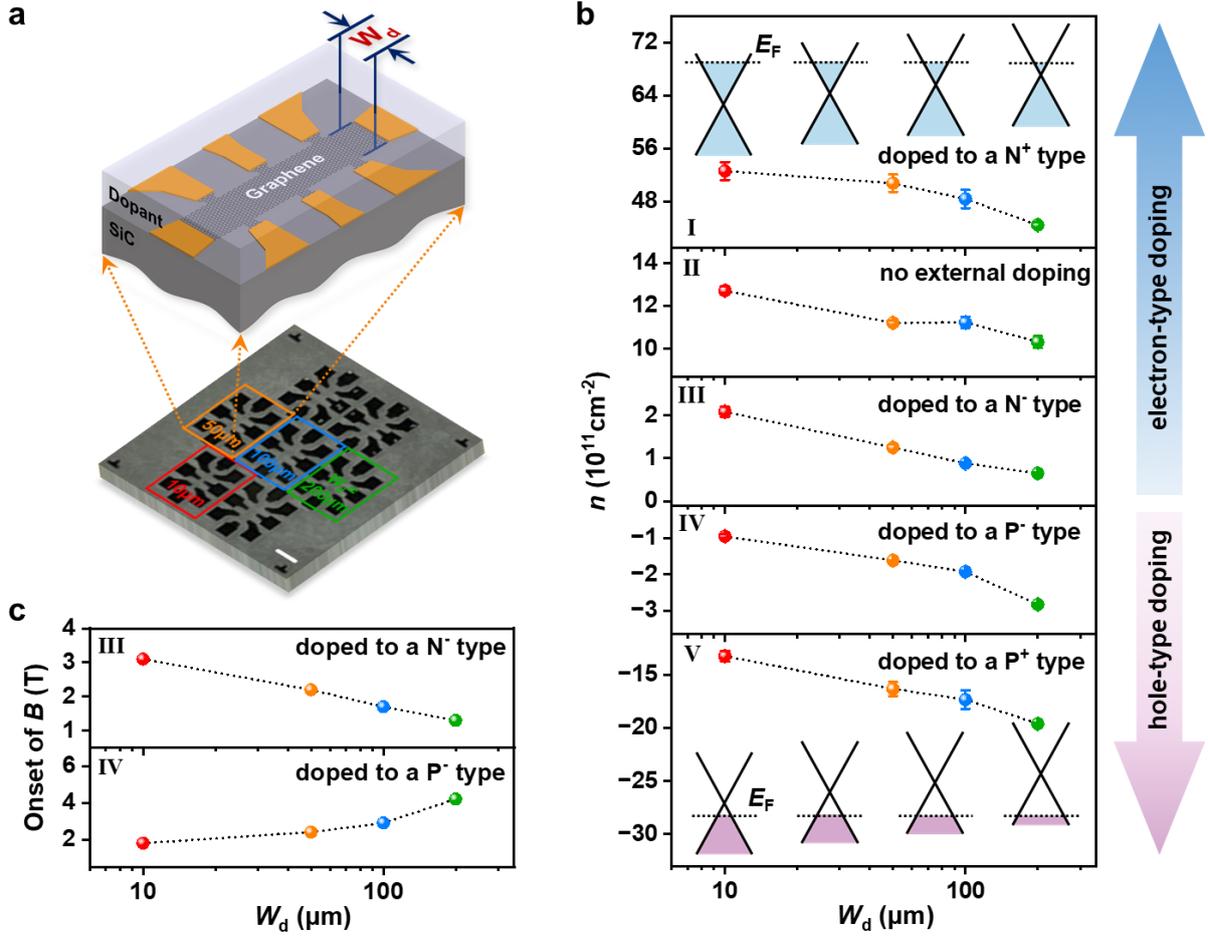

**Fig. 1.** Graphene Hall devices with different $W_d$ on the same SiC substrate and scale-dependent charge carrier density in them. (a) Photograph and schematic diagram of the graphene Hall devices with different $W_d$ on the same SiC substrate. $W_d$ varies among 10 μm (red), 50 μm (orange), 100 μm (blue) and 200 μm (green). The scale bar represents 600 μm. The schematic diagram above the photograph shows the device structure, which consists of silicon carbide, monolayer graphene, metal electrodes, and a dopant layer, from bottom to top. (b) The variation curve of carrier density ($n$) with the increase of $W_d$ under different doping conditions. The blue and red arrow represents electron-type doping and hole-type doping levels, respectively. The insets in (I) and (V) are schematic diagrams showing how $E_F$ decreases with the increase of $W_d$. (c) A plot showing the variation of the onset of magnetic field ($B$) for entering quantized Hall plateaus as a function of $W_d$ in N⁻ and P⁻ type.

To investigate whether $E_F$ continues to decrease with increasing $W_d$ beyond 200 μm, we fabricated larger graphene Hall devices with $W_d$ ranging from 400 μm to 1000 μm (Fig. S13).

It is found that for $W_d \leq 400$ μm, the carrier density and Fermi level decrease with increasing $W_d$ under N⁻ doping conditions. However, this trend becomes less pronounced for $W_d > 400$ μm.

As graphene edges can be more easily disrupted by impurities or defects, Kelvin probe force microscopy (KPFM) measurements were carried out to measure graphene samples on SiC substrates to investigate the influence from graphene edges, which are crucial for the quantum Hall effect. The results show that the doping influence on graphene's surface potential is limited to a very small scale, much smaller than 10 μm, as shown in Fig. S16, thereby ruling out the possibility that the graphene edges are the cause of the scale-dependent charge carrier density.

The phase coherence length—the distance over which an electron retains its phase—is crucial for a device's quantum behavior. When it exceeds the device's size, coherent transport leads to quantum interference; when it is shorter, these effects diminish. Typically, devices made from the same material exhibit stable phase coherence if edge scattering is negligible. Moreover, larger channel widths reduce boundary scattering and lengthen the coherence, and phase coherence generally increases with rising carrier density, regardless of doping type [31].

Since carrier density provides insights into phase coherence length, weak localization measurements were performed on graphene Hall devices with varying channel sizes at different temperatures to extract phase coherence length, as shown in Fig. S10. In electron-type doping conditions (Fig. S10a), the phase coherence length fluctuates upward with increasing channel width, indicating a decrease in carrier density. This fluctuation arises because narrower devices exhibit higher carrier density under electron-doping. In contrast, under hole-type doping (Fig. S10b), the phase coherence length consistently increases with channel width, reflecting an increase in carrier density. Notably, for $W_d \geq 400$ μm, the characteristic length of the graphene Hall device remains largely unaffected by further channel width variations (Fig. S15). This observation suggests that beyond this threshold, carrier density shows minimal dependence on device dimensions, as edge scattering can be ignored.

To determine the fundamental cause of this phenomenon of pronounced scale-dependent charge carrier density, we used transport methods, as discussed in various studies [32-34], to investigate the band structure of graphene. We aimed to explore whether the observed scale-variant phenomena are related to changes in the graphene band structure. To explore changes in the graphene band structure, we measured the longitudinal resistance ($R_{xx}$) of graphene devices with different $W_d$ at various doping conditions to extract the cyclotron mass ($m_c$) and

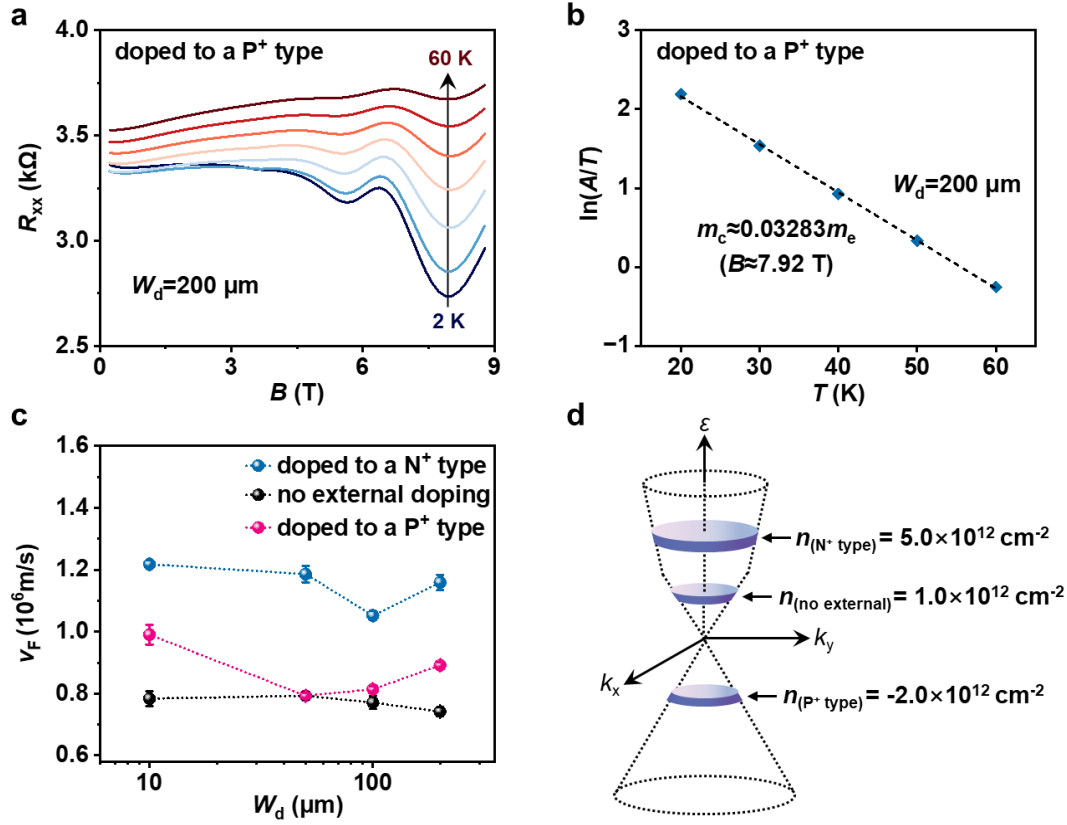

**Fig. 2.** Measurement results of Fermi velocity near the Fermi level in graphene Hall devices with different $W_d$. (a) Longitudinal resistance ($R_{xx}$) as a function of magnetic field at different temperature from 2 K (deep blue) to 60 K (dark red). The device channel width $W_d$ is 200 μm. (b) The scatter plot of ln($A/T$) extracted from the Shubnikov-de Haas (SdH) oscillation curves obtained by subtracting the background signal from the $R_{xx}$ magnetic transport curve in panel (a). The black dashed line represents the Lifshitz-Kosevich (L-K) fitting curve, yielding the cyclotron mass $m_c$. (c) Extracted Fermi velocity ($v_F$) of graphene Hall devices with different $W_d$ for N$^+$ type, P$^+$ type, and the no external doping conditions. (d) The corresponding schematic of the graphene energy band structure plotted based on the results of the Fermi velocity shown in panel (c). The black arrows represent the Fermi circle corresponding to the certain carrier density $n$.

Fermi velocity ($v_F$). The magneto-transport curves of $R_{xx}$ were obtained by applying an external magnetic field from 0 T to 9 T, as shown in Fig. 2a for a 200 μm device. As the temperature increases, the magnetic field positions corresponding to the maxima of $R_{xx}$ shift toward higher values, which is potentially attributable to the phase transition of insulator-quantum Hall conductor and thermal broadening effect in graphene [35-43]. Based on the data in Fig. 2a, the background signal was subtracted from the $R_{xx}$ curve to obtain the Shubnikov-de Haas (SdH) oscillation curves. The amplitudes ($A$) of the SdH oscillations at the magnetic field corresponding to the maximum amplitude (~7.92 T) at temperatures ranging from 20 K to 60 K were extracted and plotted in Fig. 2b. The $m_c$ was obtained by fitting the data using the Lifshitz-Kosevich (L-K) formula (1). Subsequently, the $v_F$ was calculated using formula (2), with the carrier density ($n$) at the magnetic field corresponding to the maximum amplitude and $m_c$ as inputs.

The formulas (1) and (2) are as follows [44-48]:

$$A = \frac{\lambda(T)}{\sinh(\lambda(T))} \qquad (1)$$

Where $\lambda(T) = 2\pi^2 k_B T m_c / \hbar e B$, $A$ is the amplitude of the Shubnikov-de Haas (SdH) oscillations, $k_B$ is the Boltzmann constant, $T$ is the temperature, $m_c$ is the cyclotron mass, $\hbar$ is the reduced Planck constant, $e$ is the electron charge, and $B$ is the applied magnetic field strength.

$$v_F = \frac{\hbar \sqrt{\pi n}}{m_c} \qquad (2)$$

Where $v_F$ is the Fermi velocity, $m_c$ is the cyclotron mass, and $n$ is the carrier density.

Using this transport method, the Fermi velocity for graphene Hall devices with different $W_d$ under N$^+$ type, P$^+$ type, and no external doping conditions were calculated and plotted in Fig. 2c. Each data point corresponds to the SdH measurement results shown in Fig. S17-S28. The Fermi velocities in Fig. 2c were used to estimate the energy band structure of graphene on SiC, as shown in Fig. 2d, with black arrows representing the Fermi circle for certain carrier densities. Variations in $v_F$ typically indicate slight changes in the Dirac cone's shape, with an increase in $v_F$ indicating a steeper Dirac cone in $k$-space.

The results in Fig. 2 indicate that the band structures of graphene Hall devices of varying sizes have undergone reshaping near the Fermi level, potentially leading to scale-dependent variations in carrier density. To further investigate the overall band structure, we performed angle-resolved photoemission spectroscopy (ARPES) after removing the dopant layer from the graphene Hall devices, as shown in Fig. 3. Fig. 3a presents a photograph of graphene Hall devices with $W_d$ values of 10, 50, and 200 μm. Fig. 3b displays ARPES measurement results, with the horizontal axis representing the wave vector $k$ and the vertical axis representing $E$-$E_F$.

Fig. 3c illustrates the energy difference between the Fermi level ($E_F$) and the Dirac point ($E_D$), extracted from Fig. 3b. As $W_d$ increases, the energy gap between $E_F$ and $E_D$ gradually narrows, suggesting a downward shift of $E_F$ with the increasing $W_d$. This observation is consistent with the magneto-transport measurement results shown in Fig. 1.

Fig. 3d–e show scatter plots of $E$-$E_F$ versus wave vector $k$, extracted from Fig. 3b, with Fig. 3d corresponding to the region above the Dirac point and Fig. 3e below it. The solid lines represent fitting curves using the formula $E = v_F \hbar k$ to extract the Fermi velocity ($v_F$) [49]. Fig. 3f is a scatter plot with connecting lines showing the dependence of $v_F$ on $W_d$. For electron-type doping (above the Dirac point), $v_F$ decreases as $W_d$ increases, whereas for hole-type doping (below the Dirac point), $v_F$ increases with $W_d$. These findings suggest that the band structure of graphene undergoes modifications as $W_d$ increases.

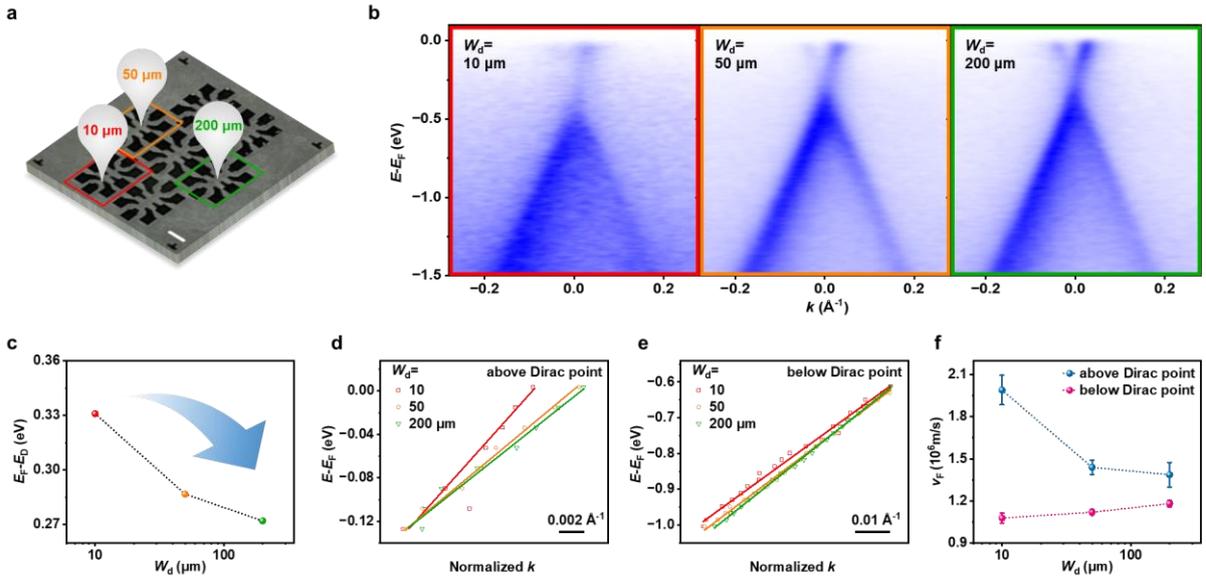

**Fig. 3.** ARPES measurement results of graphene quantum Hall devices with different $W_d$ on the same SiC substrate. (a) Photograph of graphene Hall devices with different $W_d$, where devices with $W_d$ of 10/50/200 μm are labeled in red/orange/green, respectively. The devices are doped to a N⁻ type. (b) ARPES measurement results of graphene Hall devices in panel (a). (c) Values of $E_F$-$E_D$ extracted from the ARPES measurement results in panel (b), with blue arrows indicating the downward trend. (d) Scatter plot of $E$-$E_F$ versus wave vector $k$ above the Dirac point, with a straight line representing the fitted data used to extract the Fermi velocity $v_F$. (e) Scatter plot of $E$-$E_F$ versus wave vector $k$ below the Dirac point, with a straight line representing the fitted data used to extract the Fermi velocity $v_F$ as well. (f) Scatter plot with connecting lines

of $v_F$ as a function of $W_d$ extracted from panels (d) and (e).

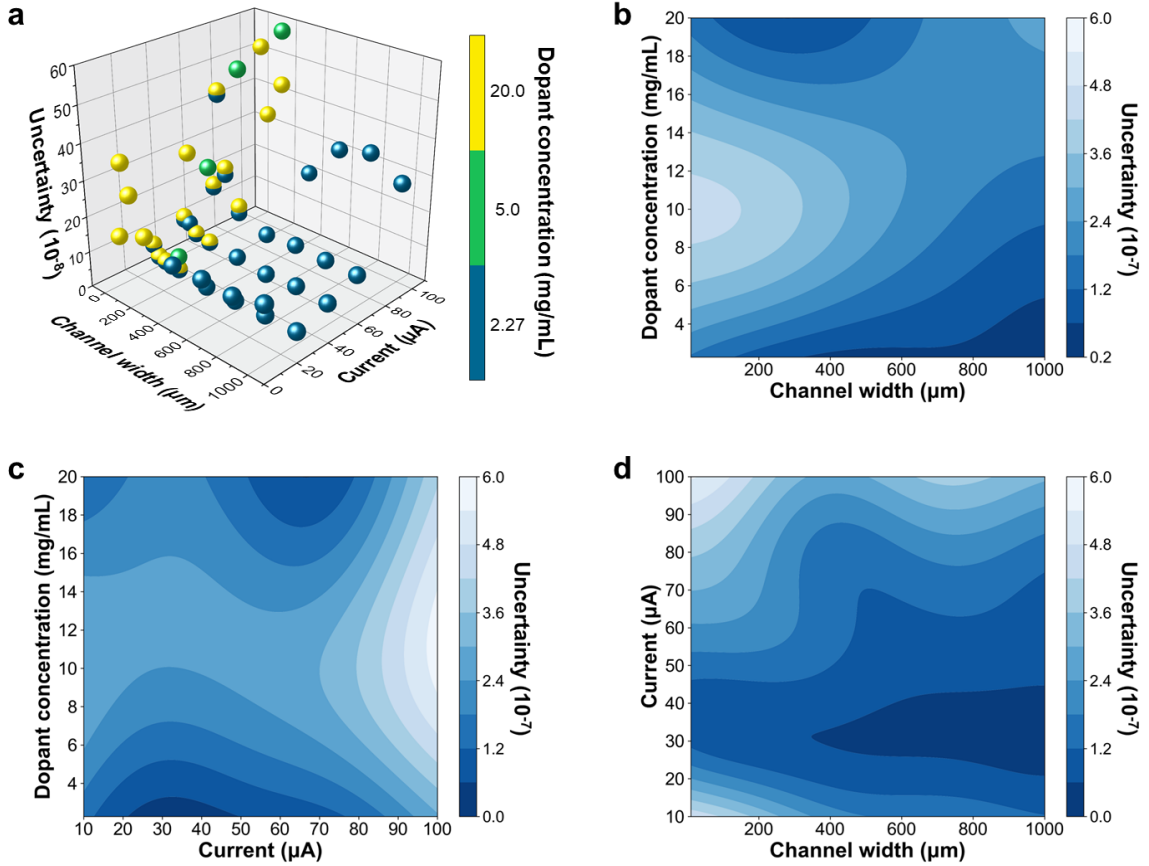

**Fig. 4.** Support Vector Regression (SVR) maps for uncertainty of graphene Hall devices. (a) 3D scatter plot for uncertainty of graphene quantum Hall devices with different conditions. (b-d) Contour plots of SVR model on two-parameter space: (b) dopant concentration vs channel width, (c) dopant concentration vs current, and (d) current vs channel width. The coded color indicates the expected uncertainty of graphene quantum Hall devices.

The measurements of the Fermi velocity ($v_F$) near the Fermi level and ARPES analyses indicate that variations in device size exert subtle yet non-negligible effects on the band structure of graphene. Furthermore, it has been suggested that electron–electron interactions in graphene enhance the dispersion velocity [50,51], thereby compressing the Dirac cone and possibly leading to a downward shift of the Fermi level with increasing device width. We propose that this band structure renormalization, driven collectively by size effects and electron–electron interactions, is the primary mechanism responsible for the observed scale-dependent variations in Fermi level and carrier density.

To minimize resistance uncertainty in graphene-based Hall devices, a systematic analysis of its dependence on both device and measurement parameters is essential. Existing studies, which often rely on the control variate method and treat parameters as independent variables, lack comprehensiveness for QHRS optimization. Machine learning has emerged as an efficient tool for uncovering hidden patterns in nonlinear datasets at lower computational cost. Fig. 4a shows a scatter plot of resistance uncertainty under varying channel widths, dopant concentrations, and probing currents. Using these data, we applied machine learning to model the relationship between uncertainty and these variables (see Experimental Section). The resulting predictions (Fig. 4b–d) indicate that the lowest uncertainty occurs at a dopant concentration of 2.27 mg/mL, a current between 27–45 μA, and a device size of 360–1000 μm. A channel width of ~360 μm was identified as optimal, offering the best trade-off between uncertainty and on-chip integration density.

These predictive results provide important guidance for the fabrication of graphene-based quantum Hall arrays (QHAs). Accordingly, we designed and fabricated a 2×3 quantum Hall resistance array with a channel width of ~360 μm, expecting it to exhibit excellent uncertainty performance. Fig. 5a shows a photograph and a schematic diagram of the home-made QHA fabricated using superconducting NbN interconnects. Fig. 5b displays the magneto-transport measurement results of the QHA under different current conditions, with the inset showing the resistance-temperature curve of the NbN superconducting electrodes. It can be observed that the QHA enters the quantized resistance plateau near ±1.5 T, with a plateau resistance value of $R_K/3$ (≈ 8.604 kΩ), at a measurement temperature of 2 K. The superconducting transition temperature of NbN remains well above 2 K under both 0 T and 9 T magnetic fields. Fig. 5c–d present the dependence of the array uncertainty on the magnetic field and probing current, respectively, as measured using a high-precision direct current comparator (DCC) bridge. As shown in Fig. 5d, the uncertainty of the QHA increases significantly when the current exceeds 80 μA. This phenomenon is primarily driven by the impact of residual system errors in the measurement equipment. The results indicate that the array achieves a minimum uncertainty of $3.0 \times 10^{-8}$ at a current of 85 μA. This demonstrates that using machine learning to guide array design and optimize resistance uncertainty is a highly promising approach worthy of further adoption.

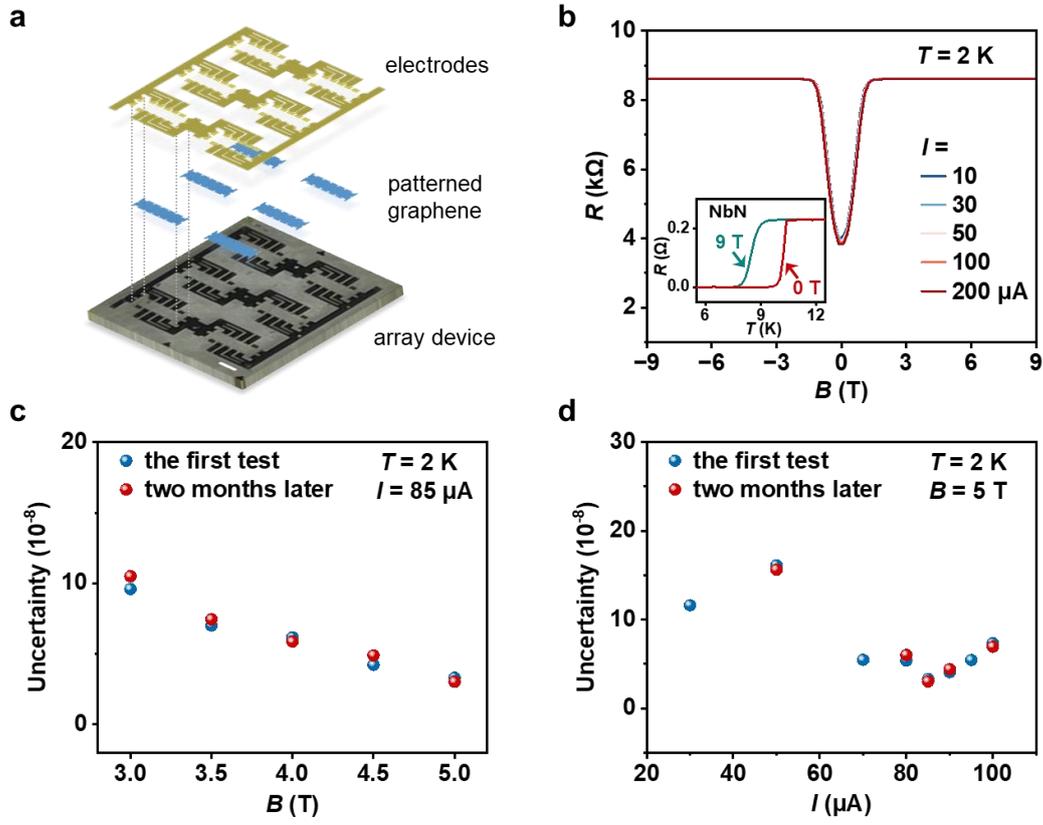

**Fig. 5.** Home-fabricated 2×3 quantum Hall array and its high-precision measurement results. (a) Photograph and schematic diagram of a 2×3 quantum Hall array. The gold-colored electrodes are Au and the superconducting NbN interconnection electrodes, while the blue region represents the patterned monolayer graphene. The photo at the bottom shows the array device. The scale bar represents 600 μm. (b) Magneto-transport curves of the QHA measured using the two-wire method at different currents. The inset shows the resistance of the NbN superconducting electrode on Au substrate as a function of temperature, measured at the magnetic field of 0 T and 9 T. (c) The scatter plot of the uncertainty of the QHA measured by high-precision DCC as a function of the magnetic field. (d) The scatter plot of the uncertainty of the QHA measured by high-precision DCC as a function of the probing current.

## 3. Conclusions

We investigate the crucial impact of device scale on graphene quantum Hall resistance standards (QHRS). Magneto-transport measurements reveal a pronounced scale-dependent carrier density: it decreases with channel width under electron doping but increases under hole doping, accompanied by a corresponding Fermi level shift. This phenomenon is attributed to subtle but significant band structure modifications, as confirmed by Fermi velocity measurements and angle-resolved photoemission spectroscopy (ARPES). Leveraging a machine learning framework under data-constrained conditions, we optimized the device geometry and identified ~360 μm as the optimal channel width, balancing minimal resistance uncertainty with high integration density. This guided the successful fabrication of an $R_K/3$ array achieving $10^{-8}$ level uncertainty. Our findings provide essential guidance for the design of high-performance, miniaturized graphene QHRS.

## 4. Experimental Section

### 4.1. The growth process of monolayer graphene for Hall device fabrication

We first perform a pre-annealing of the insulating 4H-SiC substrate in a hydrogen flow at 1400 °C to reduce the step height of the SiC substrate. Then, on the pre-annealed 4H-SiC substrate, we grow large area high-quality monolayer graphene with small step heights using chemical vapor deposition (CVD) at 1300 °C, with acetylene as the carbon source and silane as the gas catalyst. During the graphene growth process, the surface morphology of the SiC is well maintained. The ARPES image and Raman spectrum of epitaxial monolayer graphene on the silicon carbide substrate grown by this method are shown in Fig. S1.

### 4.2. The fabrication process of graphene Hall devices

We use an aluminum oxide hard mask to shield the monolayer graphene on the silicon carbide substrate during the deposition of Ti/Pd/Au metals, with thicknesses of 3/20/30 nm, respectively, to form electrical contacts. Next, maskless lithography is employed to define the graphene Hall bar pattern, followed by etching using reactive ion etching (RIE) to pattern the graphene into the Hall bar structure. Subsequently, maskless lithography is used again to define the bonding electrode pattern, and Ti/Au layers of 10/100 nm thickness are deposited. After performing the lift-off process, quantum Hall resistance devices based on the Hall bar structure are obtained. All metal deposition processes are carried out using electron beam evaporation equipment. For the maskless photolithography process, a bilayer resist system consisting of LOR-5A and S1818 is used to create an undercut.

### 4.3. The doping process of devices

We employ a spin-coating method to apply dopant layers onto the surface of graphene devices, enabling precise control over the doping level. By varying the type and concentration of the dopants, the devices can be tuned to exhibit different doping states, including $N^+$ type, $N^-$ type, $P^-$ type, and $P^+$ type. Here, the symbols '+' and '−' indicate the doping level, while 'N' and 'P' represent electron-type and hole-type doping, respectively. Specifically, $N^+$ type refers to heavily n-doped graphene with a high densty of electron carriers, whereas $N^-$ type corresponds to lightly electron-doped graphene with a lower carrier density. Similarly, $P^+$ type and $P^-$ type represent heavily and lightly hole-doped states, respectively.

To achieve $N^-$ type and $P^-$ type doping of graphene, 2,3,5,6-tetrafluoro-7,7,8,8-tetracyanoquinodimethane ($F_4TCNQ$) was used as the primary dopant. The doping solution was prepared by dissolving $F_4TCNQ$ and poly(methyl methacrylate) (PMMA) in anisole. After spin-coating and drying, the dopant film consisted solely of $F_4TCNQ$ and PMMA. By varying the concentration of $F_4TCNQ$ in the solution, different doping levels were achieved, as illustrated in Fig. S9. For $N^+$ type doping, polyethylenimine (PEI) dissolved in isopropanol was spin-coated directly onto the graphene surface. For $P^+$ type doping, a solution of 2,2'-(perfluoronaphthalene-2,6-diylidene)dimalononitrile ($F_6TCNNQ$) and PMMA dissolved in chlorobenzene was used for spin-coating. After deposition of the dopant layer, all devices were baked on a hot plate at 180 °C for 5 minutes, and further processing was conducted only after the samples cooled to room temperature. Additionally, to preserve doping efficiency and device stability, all doped samples were stored in a vacuum environment.

### 4.4. The fabrication process of graphene Hall bar arrays

Using an aluminum oxide hard mask, Ti/Pd/Au (3/20/60 nm) bonding electrodes were deposited on monolayer graphene/SiC. A Pd/Au (10/60 nm) protective layer was then applied over the channel. Uncovered graphene was removed by RIE. A 150 nm superconducting NbN electrode was deposited and patterned via maskless lithography. The channel region was then re-exposed, and the Pd/Au protective layer was etched away. Final graphene patterning was achieved using maskless lithography with a LOR-5A/S1818 bilayer resist for undercut formation. All depositions used e-beam evaporation.

### 4.5. The measurement setup

Regular and high-precision resistance measurements were performed on graphene Hall devices. Magneto-transport characterization used a 9 T Physical Property Measurement System (PPMS) with a base temperature of 2 K, integrated with a custom SR830 lock-in amplifier system.

High-precision measurements employed a Guildline 6622A direct current comparator (DCC) resistance bridge. Devices were mounted in a custom cryostat with a Gifford-McMahon refrigerator and 5 T magnet, maintaining 2 K operation monitored by an on-carrier thermometer. Measurements referenced a 10 kΩ standard resistor, with data acquired via a custom LabVIEW program.

### 4.6. Machine learning model for optimization of graphene-based Hall device

This study employs machine learning to optimize the channel width of graphene Hall devices. A Support Vector Regression (SVR) model was implemented via the sklearn.svm module in Python, using channel width, current, and dopant concentration as input features and measured uncertainty as the target. All data were standardized, and the dataset was split 80:20 for training and testing. The SVR used a Radial Basis Function (RBF) kernel with hyperparameters set as follows: penalty parameter $C = 10$ to balance model complexity and avoid overfitting; $\varepsilon = 0.001$ to ignore small errors and improve stability; and $\gamma = 0.5$ to moderate the influence of support vectors. Contour plots generated with matplotlib visualize the relationship between model predictions and input features.

**Declaration of competing interest** The authors declare that they have no known competing financial interests or personal relationships that could have appeared to influence the work reported in this paper.

### Acknowledgments

H.W. thanks Y.F. Lu (National Institute of Metrology, NIM) for his grateful help in high-precision resistance bridge measurements of the quantum Hall resistance standard at the beginning stage of this project. This work was supported by the National Key R&D Program of China (Grant No. 2023YFF0612502, 2022YFF0609800, 2022YFA1604400/03, 2021YFA1401500), National Natural Science Foundation of China (Grant Nos. 62374169,

92365204, 12274298, 62474110, 62474179, 12004406, 62074099, 12304113), Shanghai Collaborative Innovation Project (No.XTCX-KJ-2024-02), the Science and Technology Commission of Shanghai Municipality (25JD1404300), the Research Project of State Key Laboratory of Integrated Circuit Materials (No.NKLJC-Z2023-B01), the Strategic Priority Research Program of the Chinese Academy of Sciences (Grant No.XDB0670000), Shanghai Post-doctoral Excellence Program (2021515), China Postdoctoral Science Foundation (BX2021331, 2021M703338, 2021M693425, 2021K224B), ShanghaiTech Soft Matter Nanofab (SMN180827) and ShanghaiTech Material and Device Lab. Project from CETC Key Laboratory of Carbon-based Electronics (No：CKLCE0302202401).

## Authors' Contributions:

Z.K., Y.F., H.G. and R.S. contributed equally to this work. H.W. conceived and designed the research work. Y.F., C.J. and C.L performed the growth of graphene. Z.K., Y.F. and C.L. fabricated devices and arrays. Z.K. performed the low-temperature magneto-transport measurements of devices and arrays. Z.K., J.F., Y.F., Q.N. and G.A. contributed to the high-precision resistance bridge measurements of the quantum Hall resistance devices and arrays. H.G. and Z.L. performed ARPES measurements. Z.K. and Y.F. performed Raman measurements. C.J. and R.Z. performed KPFM measurements. R.S. performed the Gaussian process analysis by machine learning techniques. H.W., Z.K., Y.F., J.F., H.W., C.J., C.L and Z.L. analyzed the experimental data. H.W., Z.K. and R.S. wrote the manuscript. All the authors contributed to crucial discussions of the manuscript.

# Supplementary Material

# Pronounced scale-dependent charge carrier density in graphene quantum Hall devices

Ziqiang Kong[1,2]#, Yu Feng[1,2]#, Han Gao[3]#, Ru Sun[1,2]#, Jian Feng[5], Chengxin Jiang[1,2], Chenxi Liu[1,2], Huishan Wang[6], Yu Zhang[1,2], Junchi Song[1,2], Xuanzheng Hao[1,2], Ziceng Zhang[1,2], Yuteng Ma[7,8], Shengda Gao[9], Ren Zhu[10], Qandeel Noor[11], Ghulam Ali[11], Yumeng Yang[4], Guanghui Yu[1,2], Shujie Tang[1,2], Zhongkai Liu[3,*], Haomin Wang[1,2,*]

[1] State Key Laboratory of Materials for Integrated Circuits, Shanghai Institute of Microsystem and Information Technology, Chinese Academy of Sciences, Shanghai 200050, China

[2] Center of Materials Science and Optoelectronics Engineering, University of Chinese Academy of Sciences, Beijing 100049, China

[3] School of Physical Science and Technology, ShanghaiTech Laboratory for Topological Physics, ShanghaiTech University, Shanghai 201210, China

[4] School of Information Science and Technology, ShanghaiTech University, Shanghai 201210, China

[5] Shanghai Institute of Measurement and Testing Technology, Shanghai Key Laboratory of On-line Testing and Control Technology, Shanghai 201203, China

[6] Key Laboratory of Optoelectronic Material and Device, Department of Physics, Shanghai Normal University, Shanghai 200233, China

[7] Department of Electrical and Computer Engineering, National University of Singapore, Singapore 117583, Singapore

[8] National University of Singapore (Chong Qing) Research Institute, Chongqing Liang Jiang New Area, Chongqing 401123, China

[9] School of Advanced Technology, Xi'an Jiaotong-Liverpool University, Suzhou 215123, China

[10] Oxford Instruments, Shanghai 200030, China

[11] U.S.-Pakistan Center for Advanced Studies in Energy, National University of Sciences and Technology (NUST), H-12, Islamabad, Pakistan

*Corresponding author. E-mail: hmwang@mail.sim.ac.cn and liuzhk@shanghaitech.edu.cn

# These authors contributed equally to this work.

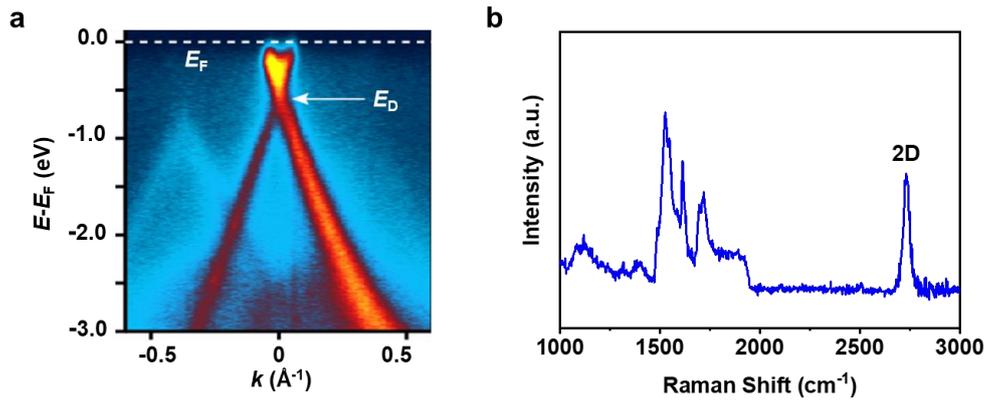

**Fig. S1.** Characterization of epitaxial graphene grown on 4H-SiC. (a) The angle-resolved photoemission spectroscopy (ARPES) intensity plot of the graphene measured around the *K* point of Brillouin zone, where $E_F$ and $E_D$ represent the Fermi level and the Dirac point energy level, respectively. (b) Raman spectra of the epitaxial monolayer graphene on 4H-SiC. The 2D peak of monolayer graphene is clearly visible.

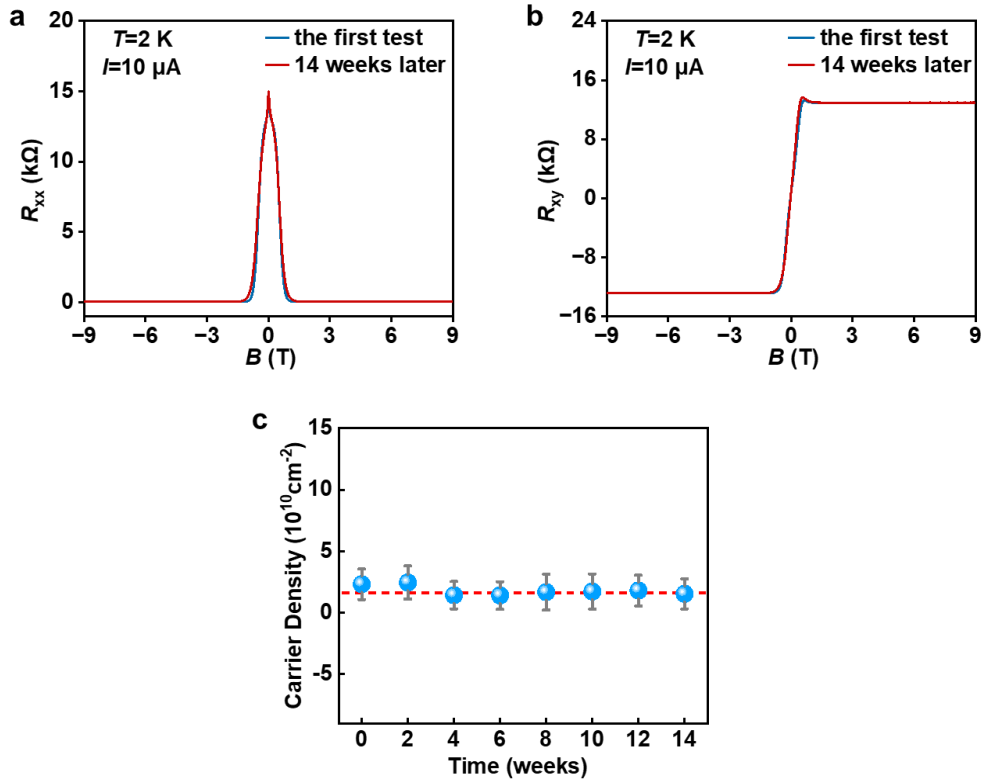

**Fig. S2.** Temporal stability of a doped graphene Hall device. (a) Magnetic field dependence of the longitudinal resistance $R_{xx}$ of the graphene Hall device. The green curve represents the result of the first measurement, and the red curve corresponds to the measurement after 14 weeks. (b) Magnetic field dependence of the Hall resistance $R_{xy}$ of the device. The Hall resistance and the longitudinal resistance were measured simultaneously. (c) Time evolution of the carrier density in the doped device over 14 weeks. It can be observed that the carrier density remains virtually constant over time, demonstrating excellent temporal stability.

The low-temperature magneto-transport measurement results of the devices with different $W_d$ in Fig. S3–S7 of the Supplementary Materials correspond to the various doping conditions among case I (doped to a $N^+$ type), II (no external doping), III (doped to a $N^-$ type), IV (doped to a $P^-$ type) and V (doped to a $P^+$ type) of Fig. 1b in the main text.

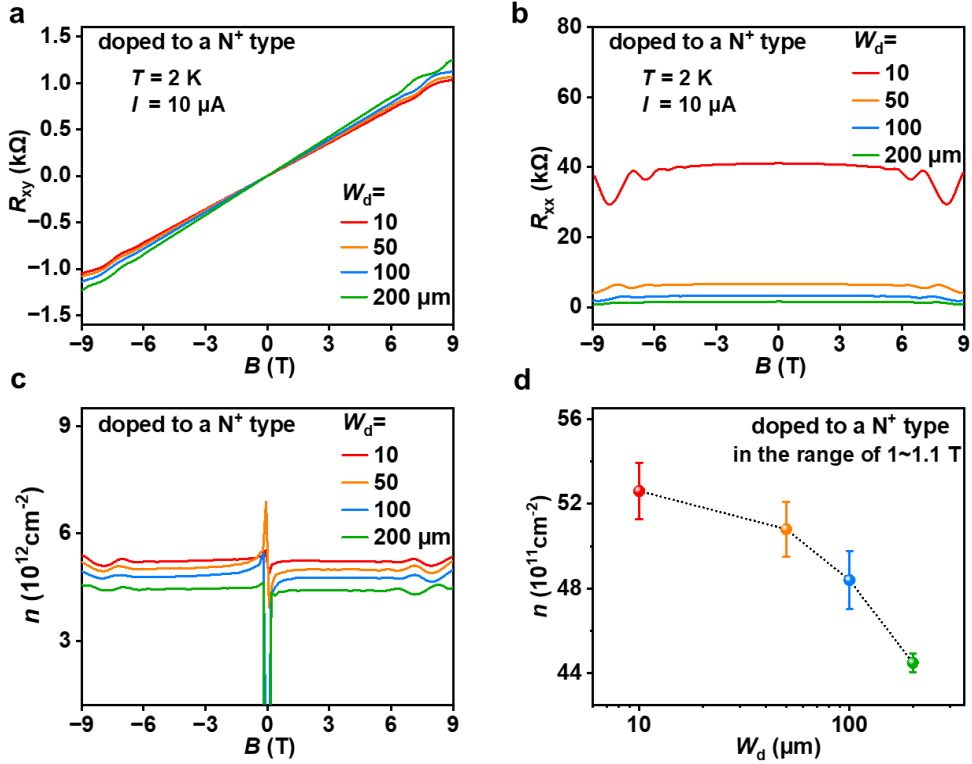

**Fig. S3.** Magneto-transport measurements of graphene Hall devices with different sizes on the same silicon carbide substrate, doped to a $N^+$ type, corresponding to case I in Fig. 1b. (a) The Hall resistance $R_{xy}$ as a function of magnetic field $B$ for graphene Hall devices with different channel widths (10 μm (red), 50 μm (orange), 100 μm (blue), and 200 μm (green)) on the same silicon carbide substrate, measured at a temperature of 2 K and a current of 10 μA. The magnetic field is scanned from -9 T to 9 T. (b) The longitudinal resistance $R_{xx}$ as a function of magnetic field $B$ for graphene Hall devices with different channel widths, with the magnetic field scanned from -9 T to 9 T. (c) The carrier density $n$ as a function of magnetic field strength for graphene Hall devices with different channel widths. (d) A scatter plot showing the extracted carrier density $n$ for graphene Hall devices with different channel widths based on panel (c), where the vertical axis values represent the average carrier density in the range of 1–1.1 T, and the standard deviation of the carrier density within this field range is used as half of the error bar.

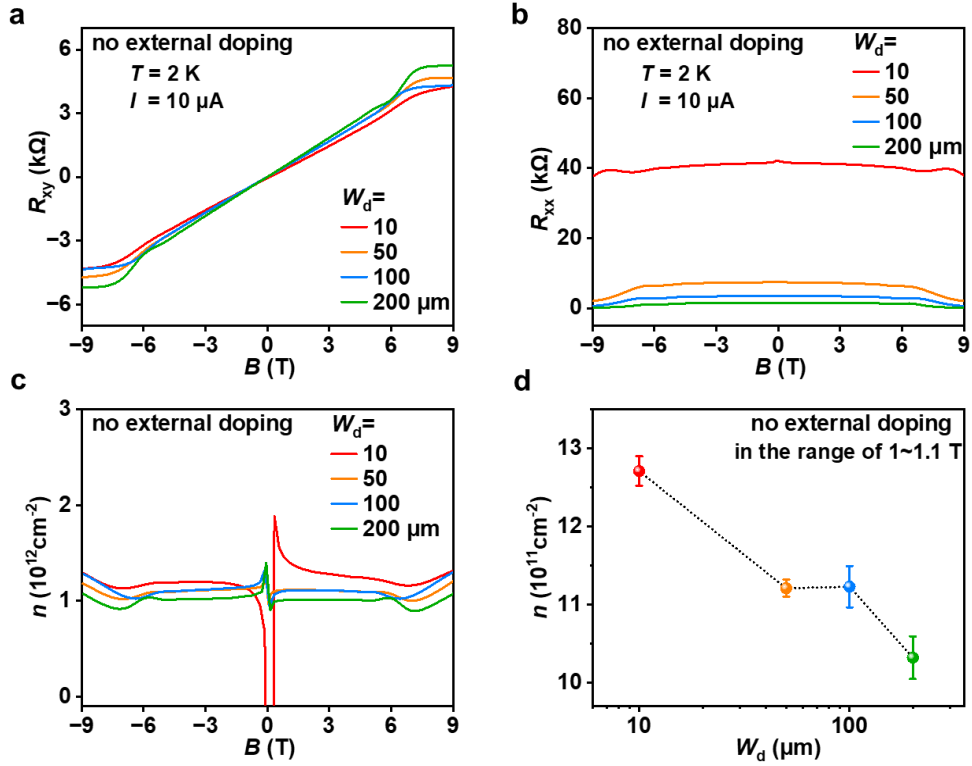

**Fig. S4.** Magneto-transport measurements of graphene Hall devices with different sizes on the same silicon carbide substrate, without external doping, corresponding to case II in Fig. 1b. (a) The Hall resistance $R_{xy}$ as a function of magnetic field $B$ for graphene Hall devices with different channel widths (10 μm (red), 50 μm (orange), 100 μm (blue), and 200 μm (green)) on the same silicon carbide substrate, measured at a temperature of 2 K and a current of 10 μA. The magnetic field is scanned from -9 T to 9 T. (b) The longitudinal resistance $R_{xx}$ as a function of magnetic field $B$ for graphene Hall devices with different channel widths, with the magnetic field scanned from -9 T to 9 T. (c) The carrier density $n$ as a function of magnetic field strength for graphene Hall devices with different channel widths. (d) A scatter plot showing the extracted carrier density $n$ for graphene Hall devices with different channel widths based on panel (c), where the vertical axis values represent the average carrier density in the range of 1–1.1 T, and the standard deviation of the carrier density within this field range is used as half of the error bar.

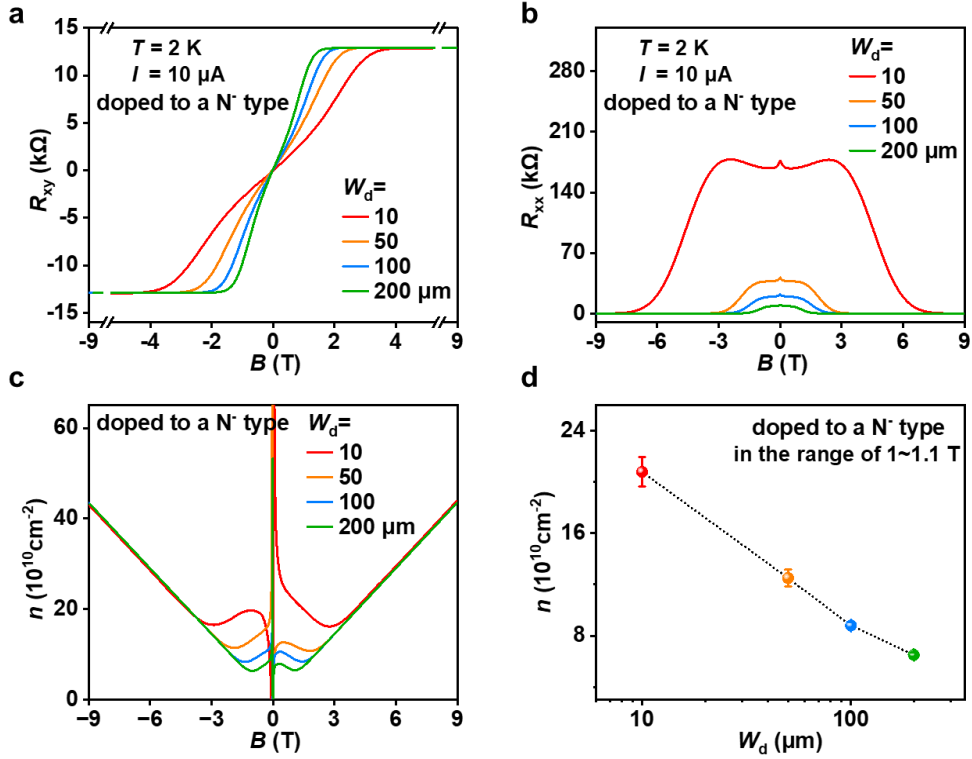

**Fig. S5.** Magneto-transport measurements of graphene Hall devices with different sizes on the same silicon carbide substrate, doped to a N⁻ type, corresponding to case III in Fig. 1b. (a) The Hall resistance $R_{xy}$ as a function of magnetic field $B$ for graphene Hall devices with different channel widths (10 μm (red), 50 μm (orange), 100 μm (blue), and 200 μm (green)) on the same silicon carbide substrate, measured at a temperature of 2 K and a current of 10 μA. The magnetic field is scanned from -9 T to 9 T. (b) The longitudinal resistance $R_{xx}$ as a function of magnetic field $B$ for graphene Hall devices with different channel widths, with the magnetic field scanned from -9 T to 9 T. (c) The carrier density $n$ as a function of magnetic field strength for graphene Hall devices with different channel widths. (d) A scatter plot showing the extracted carrier density $n$ for graphene Hall devices with different channel widths based on panel (c), where the vertical axis values represent the average carrier density in the range of 1–1.1 T, and the standard deviation of the carrier density within this field range is used as half of the error bar.

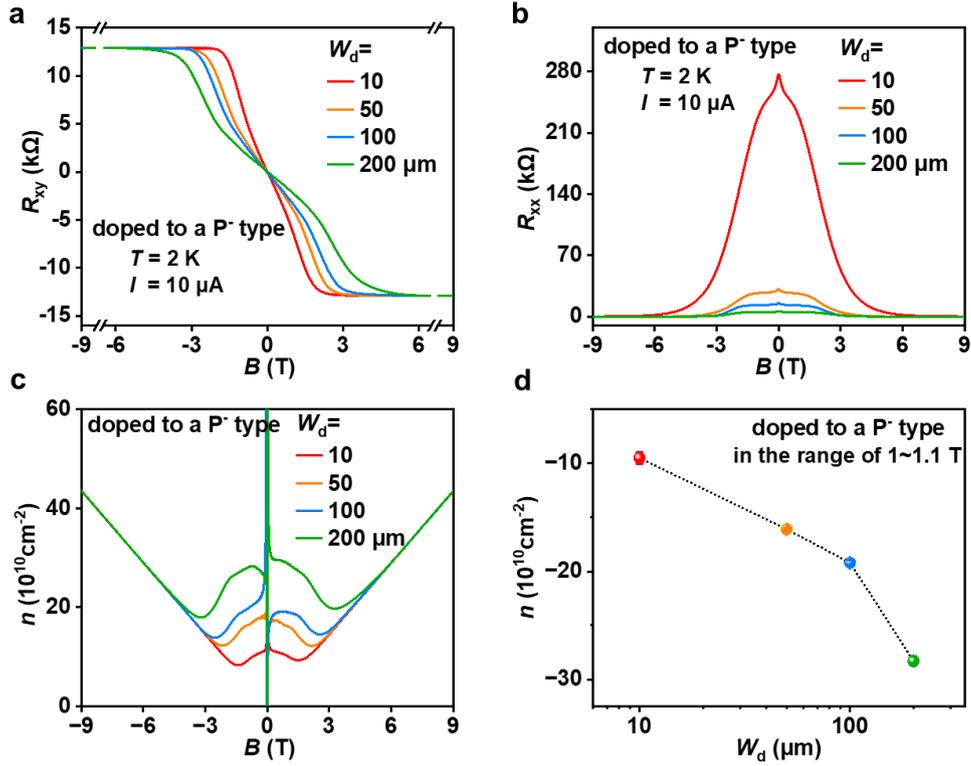

**Fig. S6.** Magneto-transport measurements of graphene Hall devices with different sizes on the same silicon carbide substrate, doped to a P⁻ type, corresponding to case IV in Fig. 1b. (a) The Hall resistance $R_{xy}$ as a function of magnetic field $B$ for graphene Hall devices with different channel widths (10 μm (red), 50 μm (orange), 100 μm (blue), and 200 μm (green)) on the same silicon carbide substrate, measured at a temperature of 2 K and a current of 10 μA. The magnetic field is scanned from -9 T to 9 T. (b) The longitudinal resistance $R_{xx}$ as a function of magnetic field $B$ for graphene Hall devices with different channel widths, with the magnetic field scanned from -9 T to 9 T. (c) The carrier density $n$ as a function of magnetic field strength for graphene Hall devices with different channel widths. (d) A scatter plot showing the extracted carrier density $n$ for graphene Hall devices with different channel widths based on panel (c), where the vertical axis values represent the average carrier density in the range of 1–1.1 T, and the standard deviation of the carrier density within this field range is used as half of the error bar.

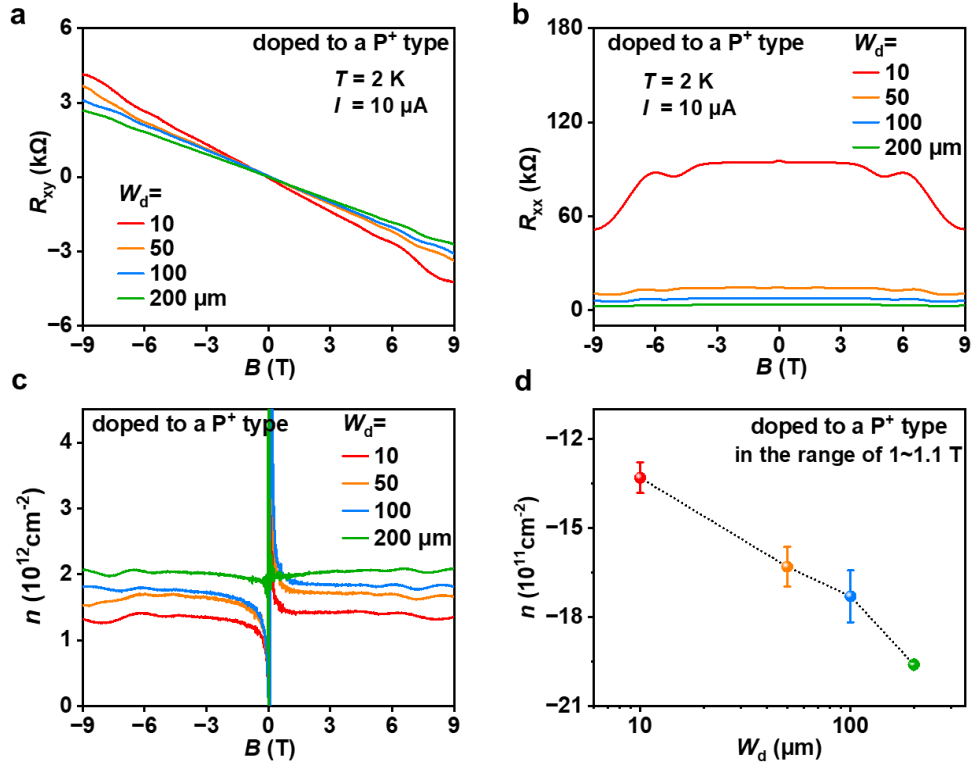

**Fig. S7.** Magneto-transport measurements of graphene Hall devices with different sizes on the same silicon carbide substrate, doped to a P$^+$ type, corresponding to case V in Fig. 1b. (a) The Hall resistance $R_{xy}$ as a function of magnetic field $B$ for graphene Hall devices with different channel widths (10 μm (red), 50 μm (orange), 100 μm (blue), and 200 μm (green)) on the same silicon carbide substrate, measured at a temperature of 2 K and a current of 10 μA. The magnetic field is scanned from -9 T to 9 T. (b) The longitudinal resistance $R_{xx}$ as a function of magnetic field $B$ for graphene Hall devices with different channel widths, with the magnetic field scanned from -9 T to 9 T. (c) The carrier density $n$ as a function of magnetic field strength for graphene Hall devices with different channel widths. (d) A scatter plot showing the extracted carrier density $n$ for graphene Hall devices with different channel widths based on panel (c), where the vertical axis values represent the average carrier density in the range of 1–1.1 T, and the standard deviation of the carrier density within this field range is used as half of the error bar.

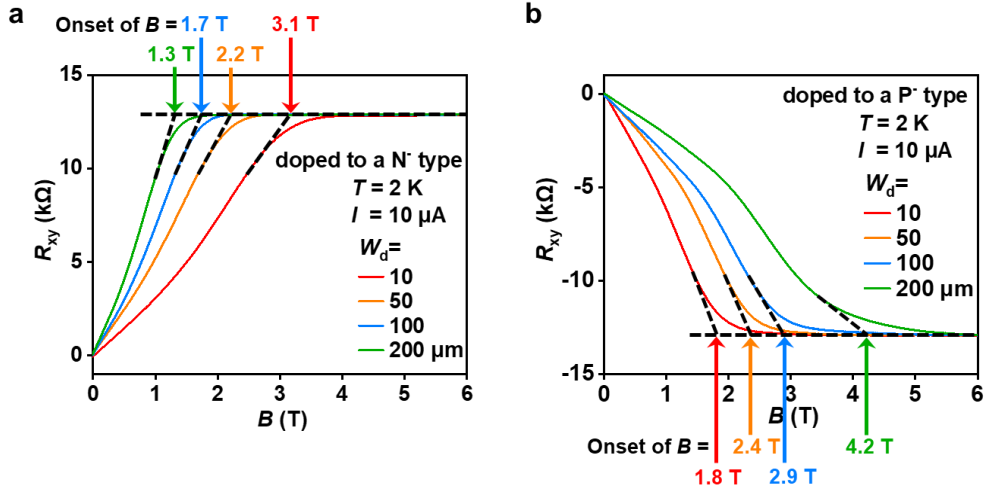

**Fig. S8.** The onset of magnetic field for entering quantized Hall plateaus as a function of $W_d$ in N⁻ and P⁻ type. (a) The Hall resistance $R_{xy}$ of the graphene devices doped to a N⁻ type as a function of magnetic field $B$, with $B$ zoomed in from 0 T to 6 T. The measurements were performed at 2 K and 10 μA. The black dashed line and colored arrows highlight the onset of $B$ at which devices with different $W_d$ enter the quantized Hall plateau. (b) The Hall resistance $R_{xy}$ of the graphene devices doped to a P⁻ type as a function of magnetic field $B$, with $B$ zoomed in from 0 T to 6 T. The black dashed line and colored arrows highlight the onset of $B$ at which devices with different $W_d$ enter the quantized Hall plateau.

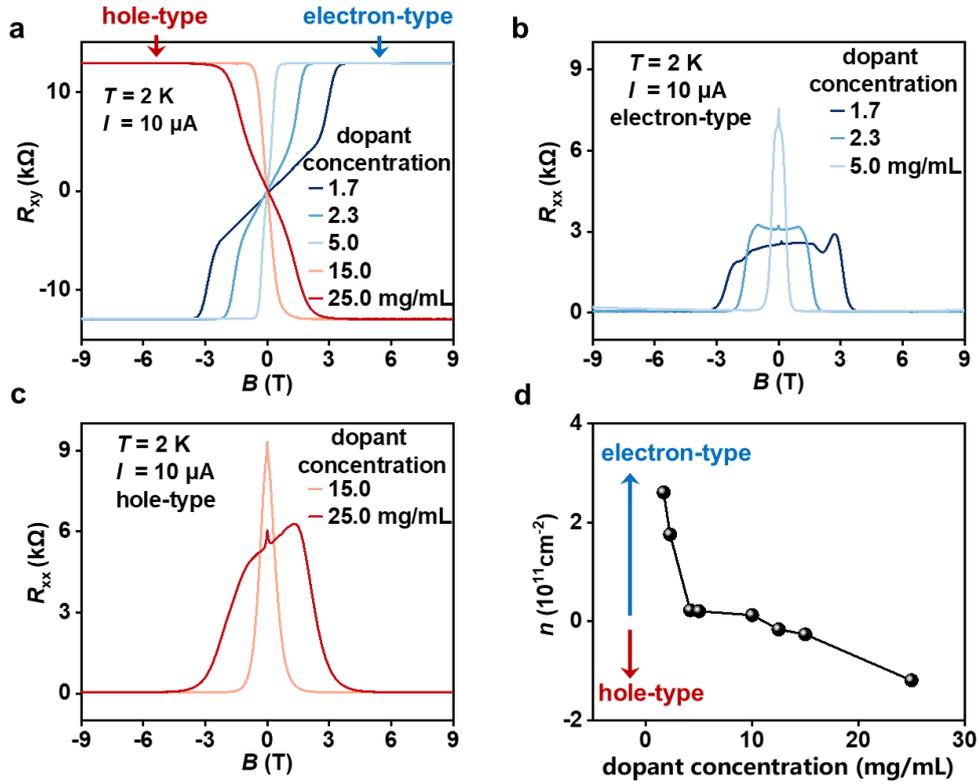

**Fig. S9.** The doping type and carrier density of graphene can be modified by varying the dopant concentration. (a) The Hall resistance $R_{xy}$ as a function of magnetic field $B$ for monolayer graphene quantum Hall standard devices on the same silicon carbide substrate at different dopant concentrations. The magnetic field is scanned from -9 T to 9 T, at a temperature of 2 K and a current of 10 µA. The dopant concentration varies from 1.7 mg/mL to 25.0 mg/mL, with the curves corresponding to electron-type and hole-type doping levels marked by blue and red arrows, respectively. (b) The longitudinal resistance $R_{xx}$ as a function of magnetic field $B$ for graphene Hall standard devices at electron-type doping levels, with dopant concentrations ranging from 1.7 to 5.0 mg/mL. The magnetic field is scanned from -9 T to 9 T. (c) The longitudinal resistance $R_{xx}$ as a function of magnetic field $B$ for graphene Hall standard devices at hole-type doping levels, with dopant concentrations ranging from 15.0 to 25.0 mg/mL. The magnetic field is scanned from -9 T to 9 T. (d) The carrier density $n$ of graphene Hall devices as a function of dopant concentration, extracted from panel (a). The blue and red arrows indicate the ranges of carrier density $n$ at electron-type and hole-type doping conditions, respectively.

It is found that by changing the concentration of the dopant, the carrier density and carrier type of the monolayer graphene quantum Hall resistance standard device on silicon carbide (SiC) could be modified. The magneto-transport results of the device at low temperature (2 K) and a current of 10 μA are shown in Fig. S9. As the dopant concentration was increased from 1.7 mg/mL to 25.0 mg/mL, the carrier type of the graphene device changed from electron-type to hole-type. The Hall resistance $R_{xy}$ magneto-transport curves for electron-type and hole-type doping are marked with blue and red arrows, respectively, as shown in Fig. S9a. Dopant concentrations from 1.7 mg/mL to 5.0 mg/mL can modify the epitaxial graphene on silicon carbide to electron-type doping, while dopant concentrations from 15.0 mg/mL to 25.0 mg/mL or even higher can tune the graphene to hole-type doping, as shown in Fig. S9b and S9c. Fig. S9b–c show the variation of $R_{xx}$ with the applied magnetic field $B$ at electron-type and hole-type doping levels, respectively. At the electron-type doping level, as the dopant concentration increases, the onset of magnetic field for entering quantized Hall plateau gradually decreases. At the hole-type doping level, as the dopant concentration increases, the onset of magnetic field for entering quantized Hall plateau gradually increases. Based on Fig. S9a, the curve of carrier density $n$ in the graphene Hall device as a function of dopant concentration can be extracted, as shown in Fig. S9d.

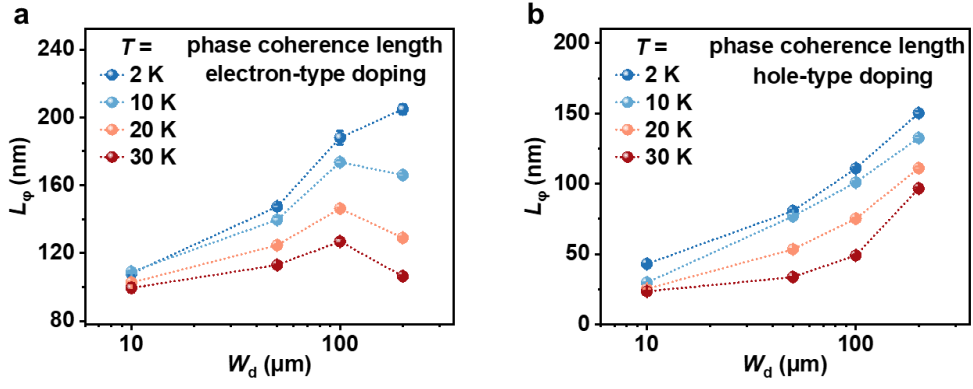

**Fig. S10.** Variation of the phase coherence length ($L_\varphi$) of graphene Hall devices with device channel width at different temperatures. The device channel widths are 10/50/100/200 μm. The measurement temperatures range from 2 K to 30 K. (a) Electron-type doping condition. (b) Hole-type doping condition.

Fig. S10 illustrates the variation of the phase coherence length ($L_\varphi$) of graphene Hall devices as a function of the device channel width at different temperatures, with the phase coherence length being extracted from the weak localization measurement data presented in Fig. S11–S12.

Fig. S11–S12 present weak localization measurement data and the corresponding fitting results of characteristic length for graphene Hall devices with different channel widths, fabricated on the same SiC substrate. Fig. S11 corresponds to electron-type doping condition, while Fig. S12 corresponds to hole-type doping condition. The device channel widths are 10/50/100/200 μm. The measurement temperatures range from 2 K to 30 K. The scatter points in the figures represent the row measurement data, and the solid lines represent the fitted curves. The characteristic lengths include the phase coherence length $L_\varphi$, the elastic intervalley scattering length $L_i$ and the elastic intravalley scattering length $L_*$.

The following are the fitting formulas:

$$\Delta\sigma = \frac{e^2}{\pi h} \times \left[ F\left(\frac{8\pi B}{\varnothing_0 L_\varphi^{-2}}\right) - F\left(\frac{8\pi B}{\varnothing_0 \{L_\varphi^{-2} + 2L_i^{-2}\}}\right) - 2F\left(\frac{8\pi B}{\varnothing_0 \{L_\varphi^{-2} + L_i^{-2} + L_*^{-2}\}}\right) \right]$$

where $F(z) = \ln z + \psi(0.5 + z^{-1})$, $\psi(x)$ is the digamma function, and $\varnothing_0 (= h/e)$ is the quantum flux. $\Delta\sigma$ represents the change in magnetoconductance: $\Delta\sigma = \sigma(B) - \sigma(B=0)$. $L_\varphi$ stands for the phase coherence length caused by inelastic scattering. Additionally, $L_i$ and $L_*$ stand for the elastic intervalley scattering length and the elastic intravalley scattering length, respectively.

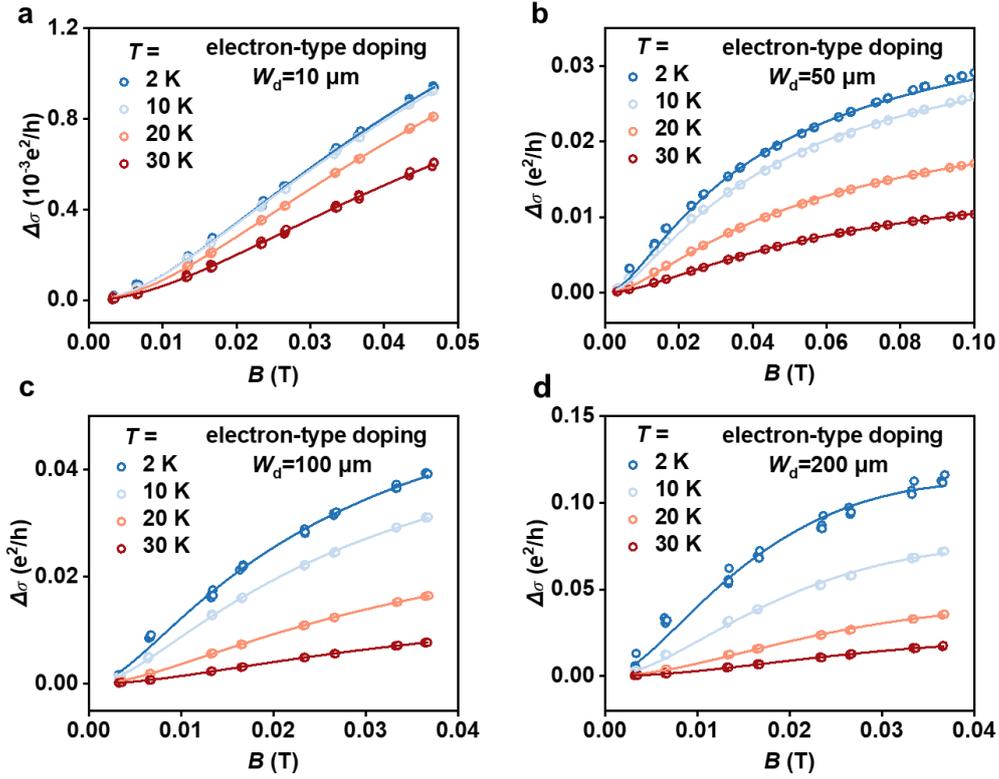

**Fig. S11.** Weak localization measurement data and fitting results of graphene Hall devices with different channel widths under electron-type doping condition. The scatter points represent the original measurement data, and the solid lines represent the fitting results. (a) The case of $W_d$=10 μm. (b) The case of $W_d$=50 μm. (c) The case of $W_d$=100 μm. (d) The case of $W_d$=200 μm.

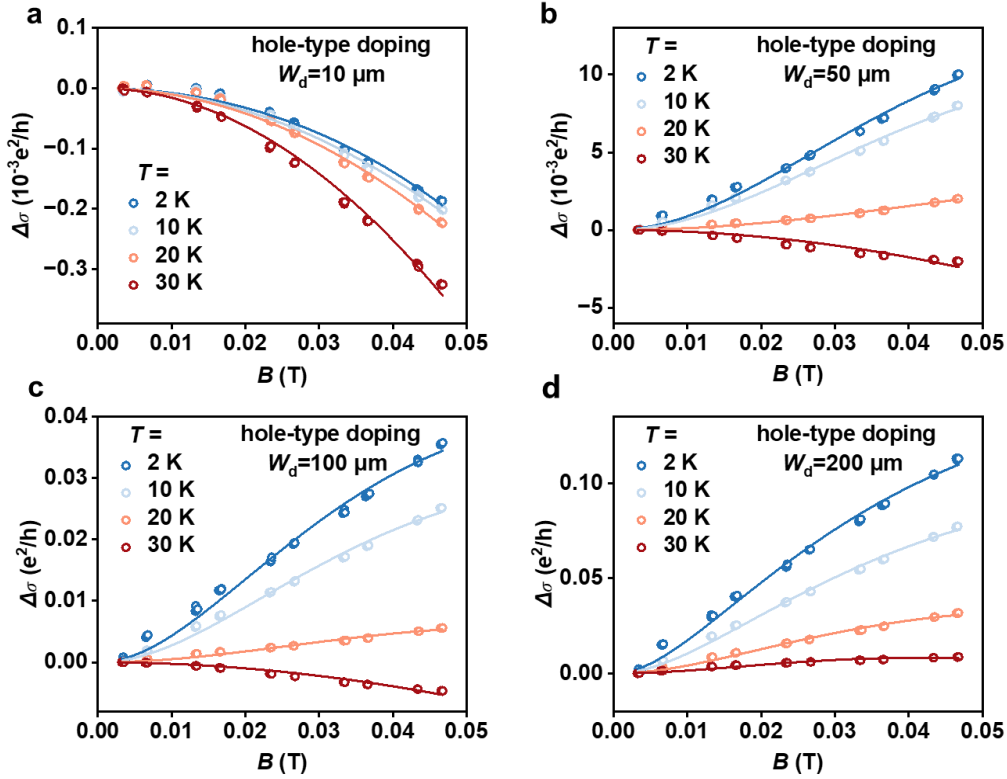

**Fig. S12.** Weak localization measurement data and fitting results of graphene Hall devices with different channel widths under hole-type doping condition. The scatter points represent the original measurement data, and the solid lines represent the fitting results. (a) The case of $W_d$=10 μm. (b) The case of $W_d$=50 μm. (c) The case of $W_d$=100 μm. (d) The case of $W_d$=200 μm.

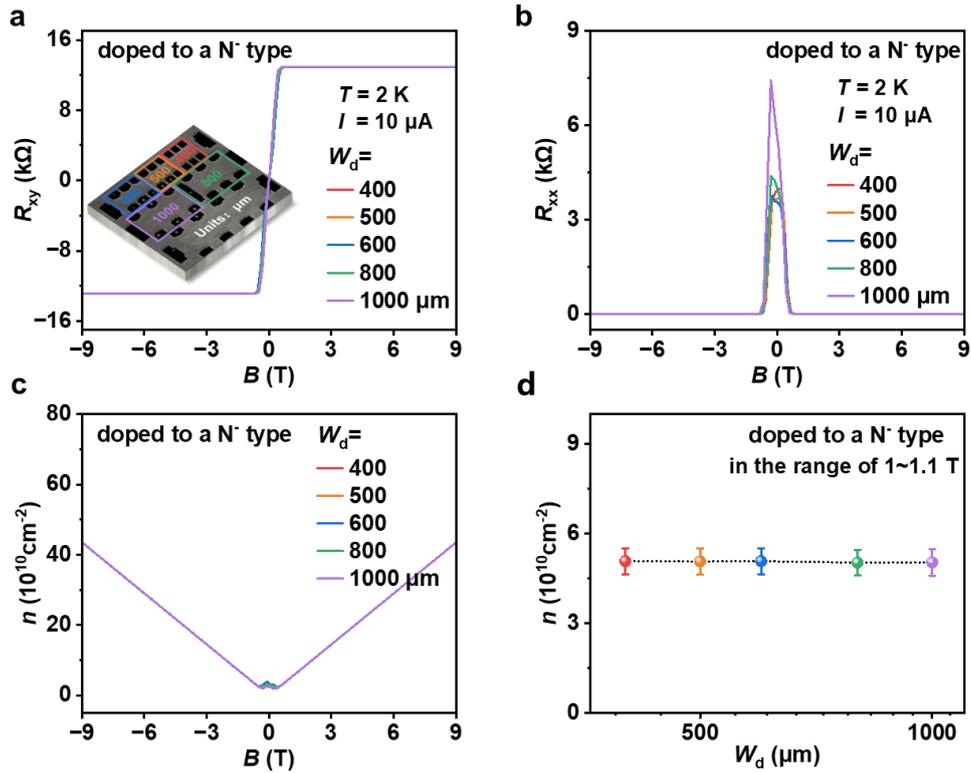

**Fig. S13.** Magneto-transport transport measurement results of larger scale graphene Hall devices on the same silicon carbide substrate doped to a N⁻ type. (a) The Hall resistance $R_{xy}$ as a function of magnetic field $B$ for graphene Hall devices with channel widths of 400 μm (red), 500 μm (orange), 600 μm (blue), 800 μm (green) and 1000 μm (purple) on the same silicon carbide substrate. The magnetic field is scanned from -9 T to 9 T. The inset shows a photograph of the large-size graphene Hall devices, with the numbers inside the box representing the channel widths in micrometers. (b) The longitudinal resistance $R_{xx}$ as a function of magnetic field $B$ for graphene Hall devices with channel widths ranging from 400 μm to 1000 μm, with the magnetic field strength scanned from -9 T to 9 T. (c) Carrier density $n$ as a function of magnetic field strength for graphene Hall devices with different channel widths. (d) A bar chart of the carrier density $n$ for graphene Hall devices with different channel widths, extracted from panel (c). The vertical axis values represents the average carrier density in the range of 1–1.1 T, with the standard deviation of the carrier density in this magnetic field range used as half of the error bar.

In order to explore whether carrier density of graphene devices continues to decrease with the increase of the device channel width when $W_d$ exceeds 200 μm, we fabricated larger scale graphene Hall devices with channel width ($W_d$) ranging from 400 μm to 1000 μm, while doping the devices to the same N⁻ doping level as shown in Fig. 1b (III). The photograph of the devices is shown in the inset of Fig. S13a, where the numbers within the boxes represent the channel width ($W_d$) in micrometers. It is found that when the channel width increased from 200 μm to 400 μm, the carrier density of the graphene device decreased from $6.5 \times 10^{10}$ cm⁻² to $5.08 \times 10^{10}$ cm⁻². However, as the channel width continued to increase from 400 μm to 1000 μm, there is no significant change in the carrier density of the graphene device, indicating that the Fermi level $E_F$ of graphene does not change. More specifically, when the channel width $W_d \geq 400$ μm, the carrier density $n$ and Fermi level $E_F$ of graphene remain almost unchanged with the increase in device size.

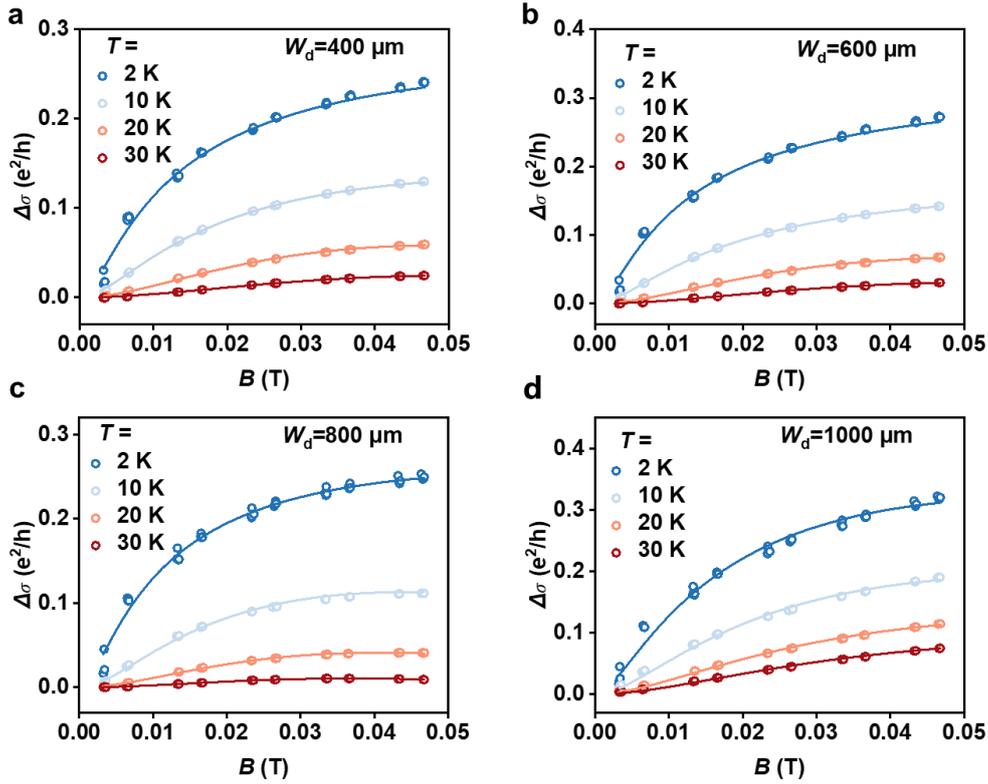

**Fig. S14.** Weak localization measurement data and fitting results of larger scale graphene Hall devices on the same SiC substrate doped to a N⁻ type. The scatter points represent the original measurement data, and the solid lines represent the fitting results. (a) The case of $W_d$=400 μm. (b) The case of $W_d$=600 μm. (c) The case of $W_d$=800 μm. (d) The case of $W_d$=1000 μm.

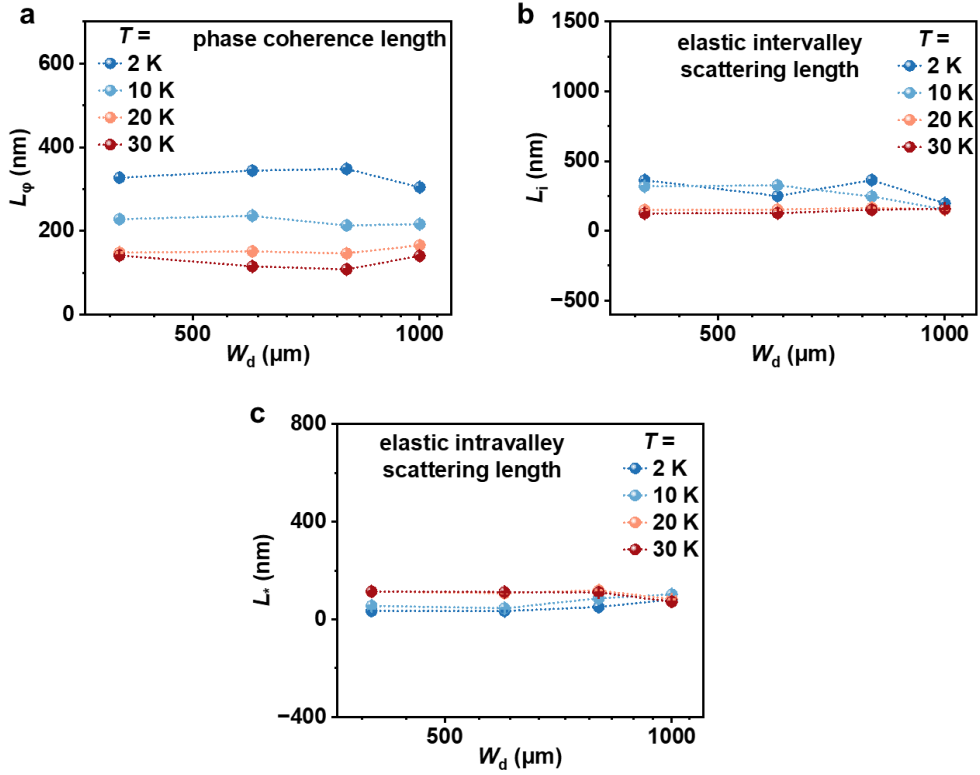

**Fig. S15.** Variation of the characteristic length of larger scale graphene Hall devices with device size at different temperatures. The device channel widths are 400/600/800/1000 μm. The measurement temperature ranges from 2 K to 30 K. (a) Phase coherence length $L_\varphi$. (b) Elastic intervalley scattering length $L_i$. (c) Elastic intravalley scattering length $L_*$.

The characteristic lengths in Fig. S15 were extracted from the weak localization source data in Fig. S14. As shown in Fig. S15, when the device channel width $W_d \geq 400$ μm, the characteristic lengths of graphene devices with varying channel widths is nearly consistent and remain within the same order of magnitude as the channel width increases from 400 μm to 1000 μm. This also suggests, from another perspective, that when $W_d \geq 400$ μm, the charge carrier density of graphene Hall devices do not exhibit significant variation with the increase of the device channel size.

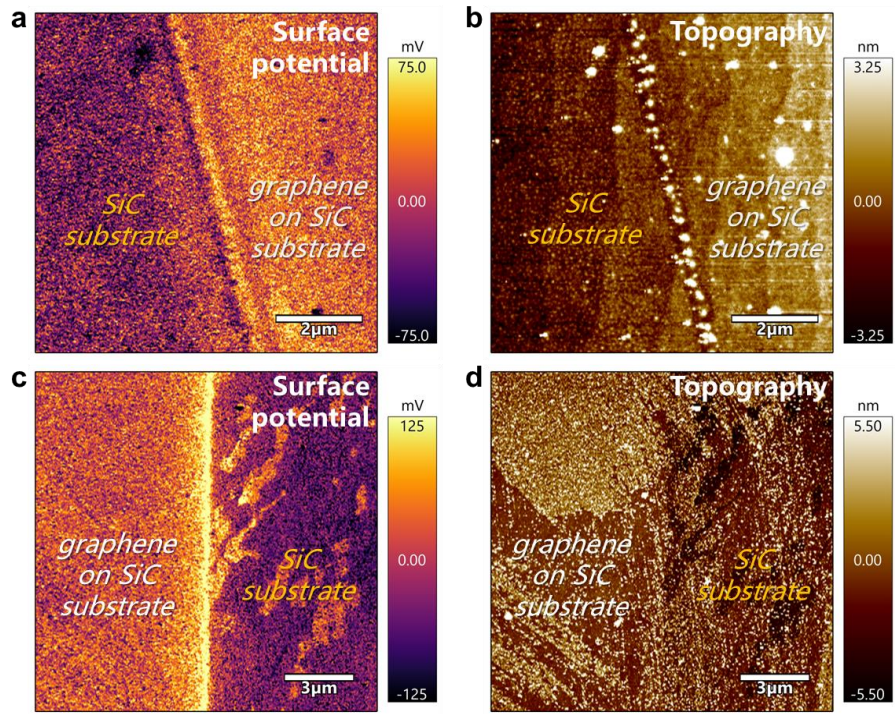

**Fig. S16.** KPFM measurement results of two graphene samples on SiC substrates. (a) Surface potential of Sample 1 measured by KPFM. The left region is the silicon carbide substrate, and the right region is the monolayer graphene on the silicon carbide substrate. The scale bar is 2 μm. (b) Topography of Sample 1. The scale bar is 2 μm. (c) Surface potential of Sample 2 measured by KPFM. The scale bar is 3 μm. The left region is monolayer graphene on the silicon carbide substrate, and the right region is the silicon carbide substrate. It can be observed that the surface potential at the graphene-SiC boundary is relatively high, but the affected width is narrow, with the width of the high-potential region being much smaller than 10 μm. (d) Topography of Sample 2. The scale bar is 3 μm.

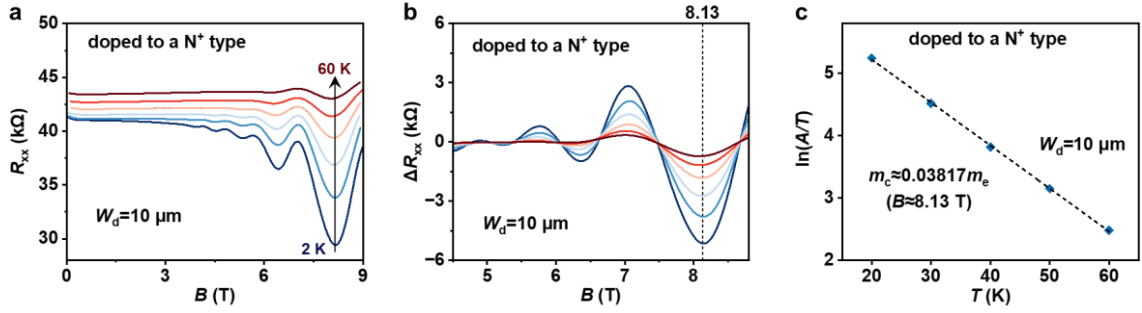

**Fig. S17.** Magnetic transport results and calculation of the cyclotron mass $m_c$ for graphene devices on a SiC substrate doped to a N$^+$ type, with a channel width $W_d$=10 μm. (a) Longitudinal resistance $R_{xx}$ of the graphene quantum Hall device as a function of the applied magnetic field $B$, with the magnetic field ranging from 0 T to 9 T. The measurement temperature varies from 2 K (dark blue) to 60 K (dark red), as indicated by the arrows. (b) Shubnikov-de Haas (SdH) oscillations extracted from the $R_{xx}$ magneto-transport curves in panel (a), after subtracting a fourth-order polynomial background signal. (c) Scatter plot of ln($A/T$) extracted from panel (b) to calculate the cyclotron mass $m_c$, where $A$ is the amplitude of the SdH oscillations and $T$ is the corresponding temperature. The fitting temperature range is from 20 K to 60 K. The black dashed line represents the Lifshitz-Kosevich fitting curve, where $m_e$ is the electron mass.

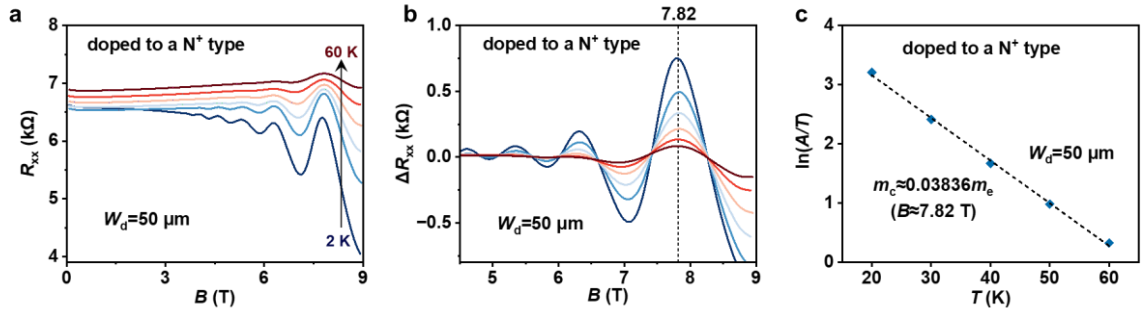

**Fig. S18.** Magnetic transport results and calculation of the cyclotron mass $m_c$ for graphene devices on a SiC substrate doped to a N$^+$ type, with a channel width $W_d$=50 μm. (a) Longitudinal resistance $R_{xx}$ of the graphene quantum Hall device as a function of the applied magnetic field $B$, with the magnetic field ranging from 0 T to 9 T. The measurement temperature varies from 2 K (dark blue) to 60 K (dark red), as indicated by the arrows. (b) Shubnikov-de Haas (SdH) oscillations extracted from the $R_{xx}$ magneto-transport curves in panel (a), after subtracting a fourth-order polynomial background signal. (c) Scatter plot of ln($A/T$) extracted from panel (b) to calculate the cyclotron mass $m_c$, where $A$ is the amplitude of the SdH oscillations and $T$ is the corresponding temperature. The fitting temperature range is from 20 K to 60 K. The black dashed line represents the Lifshitz-Kosevich fitting curve, where $m_e$ is the electron mass.

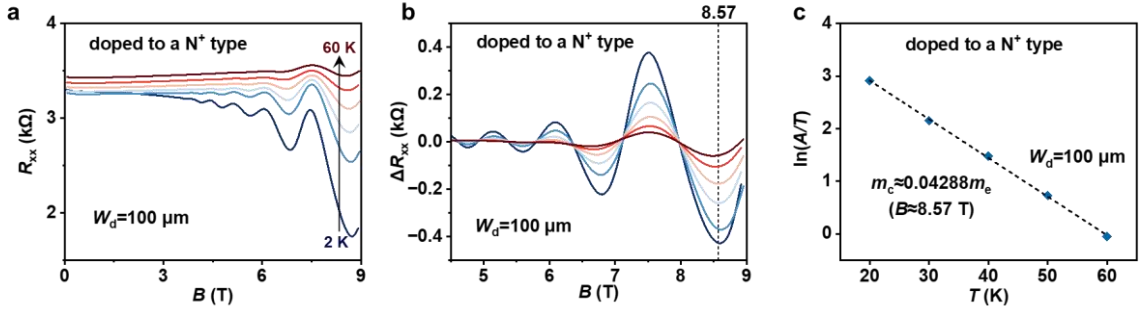

**Fig. S19.** Magnetic transport results and calculation of the cyclotron mass $m_c$ for graphene devices on a SiC substrate doped to a N$^+$ type, with a channel width $W_d$=100 μm. (a) Longitudinal resistance $R_{xx}$ of the graphene quantum Hall device as a function of the applied magnetic field $B$, with the magnetic field ranging from 0 T to 9 T. The measurement temperature varies from 2 K (dark blue) to 60 K (dark red), as indicated by the arrows. (b) Shubnikov-de Haas (SdH) oscillations extracted from the $R_{xx}$ magneto-transport curves in panel (a), after subtracting a fourth-order polynomial background signal. (c) Scatter plot of ln($A/T$) extracted from panel (b) to calculate the cyclotron mass $m_c$, where $A$ is the amplitude of the SdH oscillations and $T$ is the corresponding temperature. The fitting temperature range is from 20 K to 60 K. The black dashed line represents the Lifshitz-Kosevich fitting curve, where $m_e$ is the electron mass.

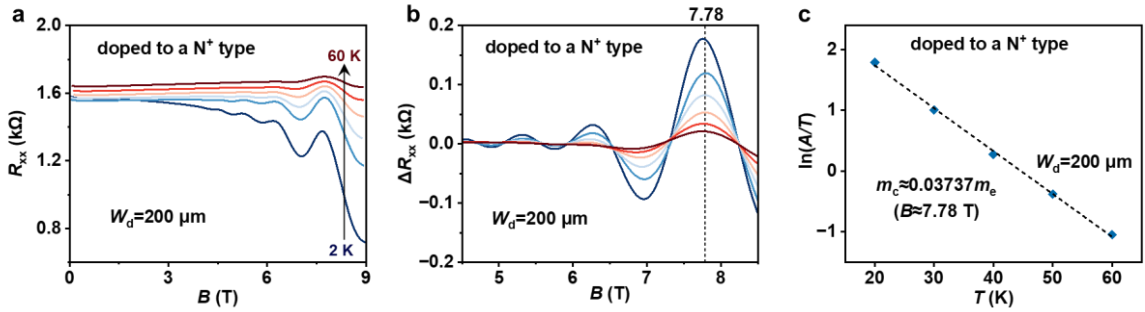

**Fig. S20.** Magnetic transport results and calculation of the cyclotron mass $m_c$ for graphene devices on a SiC substrate doped to a N$^+$ type, with a channel width $W_d$=200 μm. (a) Longitudinal resistance $R_{xx}$ of the graphene quantum Hall device as a function of the applied magnetic field $B$, with the magnetic field ranging from 0 T to 9 T. The measurement temperature varies from 2 K (dark blue) to 60 K (dark red), as indicated by the arrows. (b) Shubnikov-de Haas (SdH) oscillations extracted from the $R_{xx}$ magneto-transport curves in panel (a), after subtracting a fourth-order polynomial background signal. (c) Scatter plot of ln($A/T$) extracted from panel (b) to calculate the cyclotron mass $m_c$, where $A$ is the amplitude of the SdH oscillations and $T$ is the corresponding temperature. The fitting temperature range is from 20 K to 60 K. The black dashed line represents the Lifshitz-Kosevich fitting curve, where $m_e$ is the electron mass.

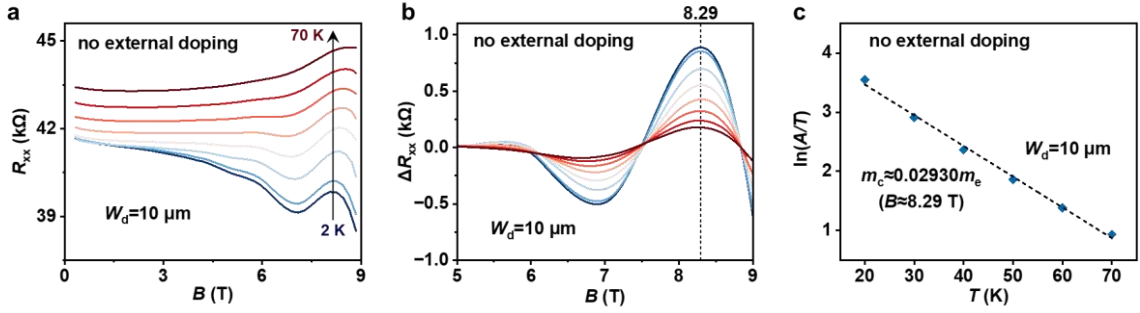

**Fig. S21.** Magnetic transport results and calculation of the cyclotron mass $m_c$ for graphene devices on a SiC substrate with no external doping, with a channel width $W_d$=10 μm. (a) Longitudinal resistance $R_{xx}$ of the graphene quantum Hall device as a function of the applied magnetic field $B$, with the magnetic field ranging from 0 T to 9 T. The measurement temperature varies from 2 K (dark blue) to 70 K (dark red), as indicated by the arrows. (b) Shubnikov-de Haas (SdH) oscillations extracted from the $R_{xx}$ magneto-transport curves in panel (a), after subtracting a fourth-order polynomial background signal. (c) Scatter plot of ln($A/T$) extracted from panel (b) to calculate the cyclotron mass $m_c$, where $A$ is the amplitude of the SdH oscillations and $T$ is the corresponding temperature. The fitting temperature range is from 20 K to 70 K. The black dashed line represents the Lifshitz-Kosevich fitting curve, where $m_e$ is the electron mass.

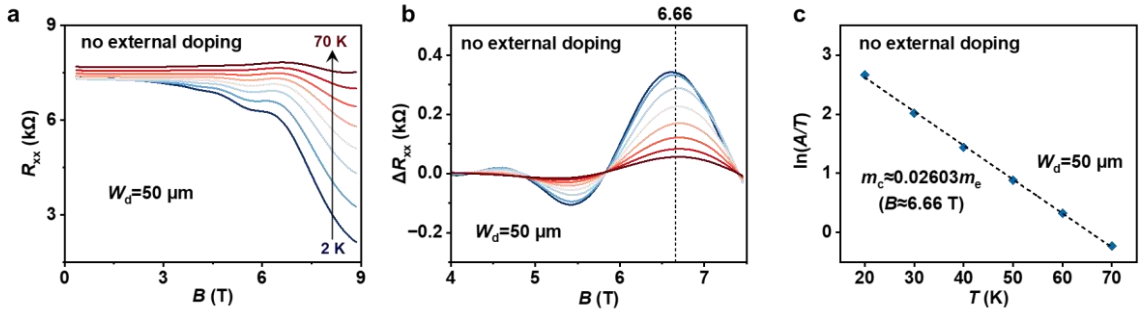

**Fig. S22.** Magnetic transport results and calculation of the cyclotron mass $m_c$ for graphene devices on a SiC substrate with no external doping, with a channel width $W_d$=50 μm. (a) Longitudinal resistance $R_{xx}$ of the graphene quantum Hall device as a function of the applied magnetic field $B$, with the magnetic field ranging from 0 T to 9 T. The measurement temperature varies from 2 K (dark blue) to 70 K (dark red), as indicated by the arrows. (b) Shubnikov-de Haas (SdH) oscillations extracted from the $R_{xx}$ magneto-transport curves in panel (a), after subtracting a fourth-order polynomial background signal. (c) Scatter plot of ln($A/T$) extracted from panel (b) to calculate the cyclotron mass $m_c$, where $A$ is the amplitude of the SdH oscillations and $T$ is the corresponding temperature. The fitting temperature range is from 20 K to 70 K. The black dashed line represents the Lifshitz-Kosevich fitting curve, where $m_e$ is the electron mass.

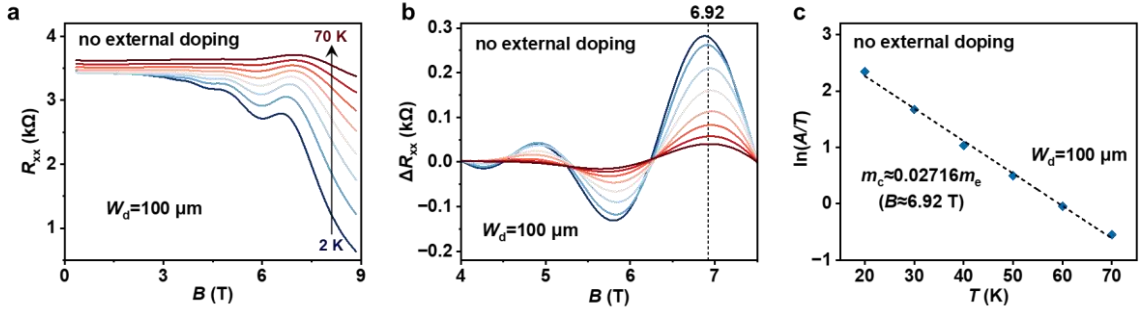

**Fig. S23.** Magnetic transport results and calculation of the cyclotron mass $m_c$ for graphene devices on a SiC substrate with no external doping, with a channel width $W_d$=100 μm. (a) Longitudinal resistance $R_{xx}$ of the graphene quantum Hall device as a function of the applied magnetic field $B$, with the magnetic field ranging from 0 T to 9 T. The measurement temperature varies from 2 K (dark blue) to 70 K (dark red), as indicated by the arrows. (b) Shubnikov-de Haas (SdH) oscillations extracted from the $R_{xx}$ magneto-transport curves in panel (a), after subtracting a fourth-order polynomial background signal. (c) Scatter plot of ln($A/T$) extracted from panel (b) to calculate the cyclotron mass $m_c$, where $A$ is the amplitude of the SdH oscillations and $T$ is the corresponding temperature. The fitting temperature range is from 20 K to 70 K. The black dashed line represents the Lifshitz-Kosevich fitting curve, where $m_e$ is the electron mass.

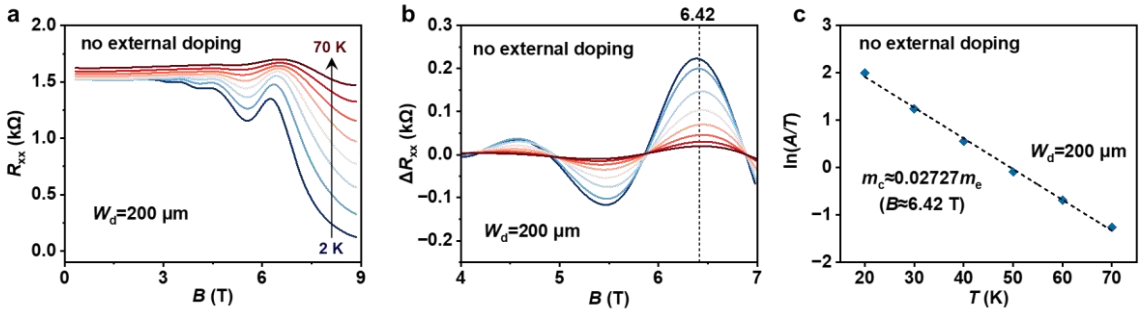

**Fig. S24.** Magnetic transport results and calculation of the cyclotron mass $m_c$ for graphene devices on a SiC substrate with no external doping, with a channel width $W_d$=200 μm. (a) Longitudinal resistance $R_{xx}$ of the graphene quantum Hall device as a function of the applied magnetic field $B$, with the magnetic field ranging from 0 T to 9 T. The measurement temperature varies from 2 K (dark blue) to 70 K (dark red), as indicated by the arrows. (b) Shubnikov-de Haas (SdH) oscillations extracted from the $R_{xx}$ magneto-transport curves in panel (a), after subtracting a fourth-order polynomial background signal. (c) Scatter plot of ln($A/T$) extracted from panel (b) to calculate the cyclotron mass $m_c$, where $A$ is the amplitude of the SdH oscillations and $T$ is the corresponding temperature. The fitting temperature range is from 20 K to 70 K. The black dashed line represents the Lifshitz-Kosevich fitting curve, where $m_e$ is the electron mass.

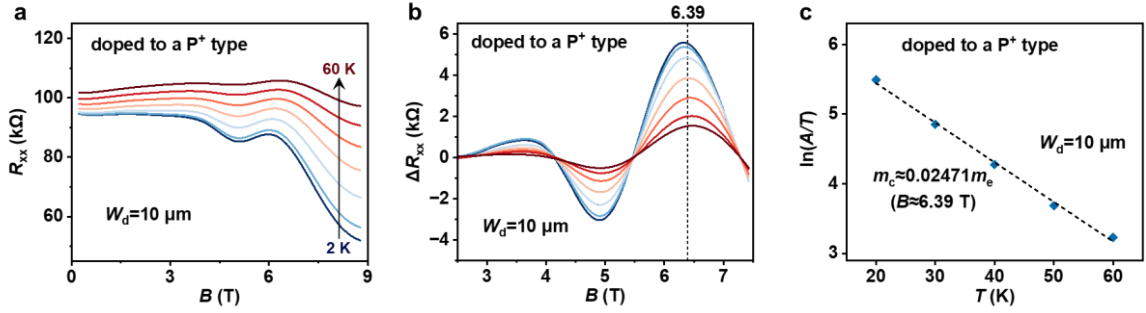

**Fig. S25.** Magnetic transport results and calculation of the cyclotron mass $m_c$ for graphene devices on a SiC substrate doped to a P$^+$ type, with a channel width $W_d$=10 μm. (a) Longitudinal resistance $R_{xx}$ of the graphene quantum Hall device as a function of the applied magnetic field $B$, with the magnetic field ranging from 0 T to 9 T. The measurement temperature varies from 2 K (dark blue) to 60 K (dark red), as indicated by the arrows. (b) Shubnikov-de Haas (SdH) oscillations extracted from the $R_{xx}$ magneto-transport curves in panel (a), after subtracting a fourth-order polynomial background signal. (c) Scatter plot of ln($A/T$) extracted from panel (b) to calculate the cyclotron mass $m_c$, where $A$ is the amplitude of the SdH oscillations and $T$ is the corresponding temperature. The fitting temperature range is from 20 K to 60 K. The black dashed line represents the Lifshitz-Kosevich fitting curve, where $m_e$ is the electron mass.

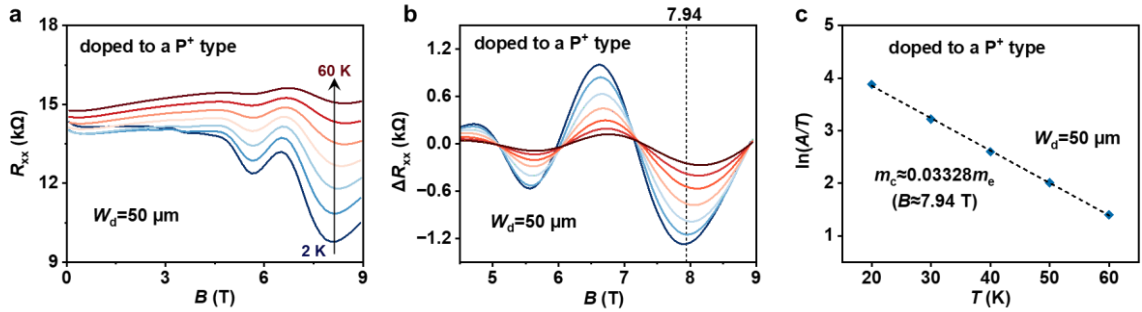

**Fig. S26.** Magnetic transport results and calculation of the cyclotron mass $m_c$ for graphene devices on a SiC substrate doped to a P$^+$ type, with a channel width $W_d$=50 μm. (a) Longitudinal resistance $R_{xx}$ of the graphene quantum Hall device as a function of the applied magnetic field $B$, with the magnetic field ranging from 0 T to 9 T. The measurement temperature varies from 2 K (dark blue) to 60 K (dark red), as indicated by the arrows. (b) Shubnikov-de Haas (SdH) oscillations extracted from the $R_{xx}$ magneto-transport curves in panel (a), after subtracting a fourth-order polynomial background signal. (c) Scatter plot of ln($A/T$) extracted from panel (b) to calculate the cyclotron mass $m_c$, where $A$ is the amplitude of the SdH oscillations and $T$ is the corresponding temperature. The fitting temperature range is from 20 K to 60 K. The black dashed line represents the Lifshitz-Kosevich fitting curve, where $m_e$ is the electron mass.

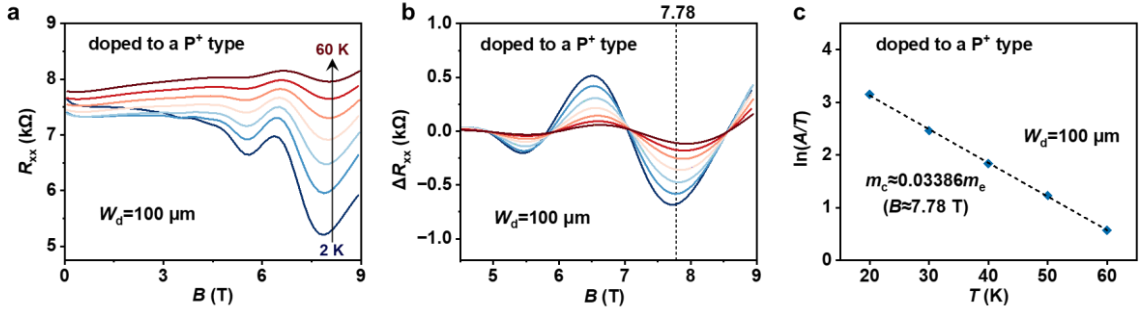

**Fig. S27.** Magnetic transport results and calculation of the cyclotron mass $m_c$ for graphene devices on a SiC substrate doped to a P$^+$ type, with a channel width $W_d$=100 μm. (a) Longitudinal resistance $R_{xx}$ of the graphene quantum Hall device as a function of the applied magnetic field $B$, with the magnetic field ranging from 0 T to 9 T. The measurement temperature varies from 2 K (dark blue) to 60 K (dark red), as indicated by the arrows. (b) Shubnikov-de Haas (SdH) oscillations extracted from the $R_{xx}$ magneto-transport curves in panel (a), after subtracting a fourth-order polynomial background signal. (c) Scatter plot of ln($A/T$) extracted from panel (b) to calculate the cyclotron mass $m_c$, where $A$ is the amplitude of the SdH oscillations and $T$ is the corresponding temperature. The fitting temperature range is from 20 K to 60 K. The black dashed line represents the Lifshitz-Kosevich fitting curve, where $m_e$ is the electron mass.

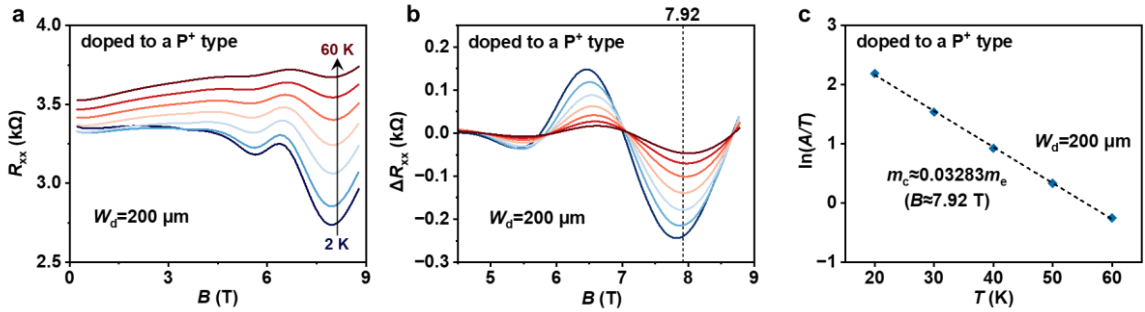

**Fig. S28.** Magnetic transport results and calculation of the cyclotron mass $m_c$ for graphene devices on a SiC substrate doped to a P$^+$ type, with a channel width $W_d$=200 μm. (a) Longitudinal resistance $R_{xx}$ of the graphene quantum Hall device as a function of the applied magnetic field $B$, with the magnetic field ranging from 0 T to 9 T. The measurement temperature varies from 2 K (dark blue) to 60 K (dark red), as indicated by the arrows. (b) Shubnikov-de Haas (SdH) oscillations extracted from the $R_{xx}$ magneto-transport curves in panel (a), after subtracting a fourth-order polynomial background signal. (c) Scatter plot of ln($A/T$) extracted from panel (b) to calculate the cyclotron mass $m_c$, where $A$ is the amplitude of the SdH oscillations and $T$ is the corresponding temperature. The fitting temperature range is from 20 K to 60 K. The black dashed line represents the Lifshitz-Kosevich fitting curve, where $m_e$ is the electron mass.